\documentclass[a4paper,12pt,twoside]{book}
\usepackage[T1]{fontenc}		
\usepackage[utf8]{inputenc}		
\usepackage[italian, english]{babel}		
\usepackage{layaureo}			
\usepackage{url}				
\usepackage{amsmath,amssymb}	
\usepackage{graphicx}
\usepackage{textcomp}
\usepackage{comment}
\usepackage{geometry}

\begin{document}

\author{Gentian Kasa}
\title{Hypercomputation: Towards an extension of the classical notion of Computability?}
\date{}
\maketitle

\frontmatter

\tableofcontents

\chapter{}

\begin{center}
This work is dedicated to my family:\\ \emph{Qemal, Makbule} and \emph{Klodian Kasa}\\ who have supported me in the writting of this thesis.\\ I can not think of any words to describe \\ my gratitude towards them.
\end{center}

\mainmatter
\chapter{Classical Calculability.}

In this chapter we'll introduce the classical model of computability with some of its limitations. To complete such a task successfully is needed a brief description of the \emph{Turing Machine} (we'll call it \emph{TM} from now on) and a specification of its power. Following that, we'll introduce some variations of this model and their qualities. In the last part of the chapter we'll analize the phisical steps of the TM and the phisical limits imposed on them.

\section{The Turing Machine.}

The TM is an ideal machine which, given a well defined set of rules, manipulates\footnote{Reads or writes via a \emph{scanning head}.} data contained in an infinite \emph{tape}.\footnote{In some cases it is said that the tape must not be limited; in other words, is required the possibility of adding more tape if it might be necessary.} The tape can be seen as a sequence of squares in each of which is possible to write a symbol $a_i$ taken from a finite predetermined \emph{alphabet} $A$. In every moment in time the TM is in a determined \emph{state of mind} $s_i$ taken from a finite predetermined set of states $S$.

Formally, one can say that a TM $T$ is a structure of the form:
\[ T = \left\langle S, s_0, F, A, \delta \right\rangle \]
where:
\begin{itemize}
\item $S$ is the set of the possible states of mind of the TM.
\item $s_0 \in S$ is the \emph{initial state} of the TM.
\item $F \subseteq S$ is the set of the \emph{final states} of the TM.
\item $A$ is the alphabet used by the TM.
\item $\delta: S \times A \to S \times A \times \left\{l, n, r \right\}\footnote{This set indicates the movements of the scanning head: left, no movement, right.}$ is the \emph{transition function} of the TM.
\end{itemize}

The instructions of the machine are quintuples like the following:
\[ \left\langle s_i, \alpha_i, s_j, \alpha_j, m \right\rangle \]
which are equivalent to:
\[ \delta(s_i, \alpha_i) = \left\langle s_j, \alpha_j, m \right\rangle \]

For more information on the subject the interested reader could consult \cite{sep:tm}.

\section{The Church-Turing Thesis.}

In this section we'll make an analysis of the \emph{Turing Thesis}. To do so, we'll make a detailed analysis of the \S9 of the paper of Turing, \cite{turing:machine}.

The TM is, basically, an abstraction of the operations made by humans who calculate. From the description given to the TM, one is sure that it is a not intelligent mechanical object. As we will see later, the elementary operations of the TM can hardly be further simplified. The real problem is the completeness of the formalization, as one can understand from the following quote:
\begin{quote}
No attempt has yet been made to show that the ‘computable’ numbers include all numbers which would naturally be regarded as computable. All arguments which can be given are bound to be, fundamentally, appeals to intuition, and for this reason rather unsatisfactory mathematically. The real question at issue is "What are the possible processes which can be carried out in computing a number?" \cite{turing:machine}
\end{quote}

Each formal description proposed for the class of the effectively computable functions with mechanical means comes with the problem of establishing not only its adequacy, but also the completeness of the description towards a non formally described set. Does the TM compute all the numbers that are naturally computable? Given a number, what are the possible operations to manipulate it? 

Demonstrating that a TM computes all the numbers that are naturally computable is a problem on which, as Turing said in the aforementioned quote, all arguments which can be given are bound to be appeals to intuition, making them rather unsatisfactory mathematically. That's why we pay attention to the meccanical, effective, operations needed for the computation of a number.

\begin{quote}
The arguments which I shall use are of three kinds.

	\begin{enumerate}
	\item A direct appeal to intuition.
	\item A proof of the equivalence of two definitions (in case the new definition has a greater intuitive appeal).
	\item Giving examples of large classes of numbers which are computable. \\
	\end{enumerate}

Once it is granted that computable numbers are all ``computable'' several other propositions of the same character follow. In particular, it follows that, if there is a general process for determining whether a formula of the Hilbert function calculus is provable, then the determination can be carried out by a machine. \cite{turing:machine}
\end{quote}

From this quote, one can perceive by intuition where Turing is trying to go. Being the TM a mechanical object which executes atomical operations\footnote{Operations which are not furtherly semplifiable.} and succeeding in producing arguments in favor of the fact that all the naturally computable numbers are computable in a mechanical way by a human being, we can deduce several things, for example that if one can prove that a function is calculable, if there is a process during which it is calculated, this process can be executed in a mechanical way by a TM. By contrast, if it can be established that a TM can not solve a given problem then neither a human being operating in a mechanical manner can solve it.

Turing, as noticed by Sieg in \cite{sieg:computability}, analyzed the human being that was calculating mechanically. Once the steps of the calculations were identified, Turing (in \cite{turing:machine}) made them atomical so that it was possible an infinity of combinations of them to make an infinite amount of calculations.

The necessity of the atomicity of the possible operations that the machine can make is explained by Turing himself in the following quote:

\begin{quote}
I. [Type (a)]. [ \dots ]
Computing is normally done by writing certain symbols on paper. We may suppose this paper is divided into squares like a child's arithmetic book. In elementary arithmetic the two-dimensional character of the paper is sometimes used. But such a use is always avoidable, and I think that it will be agreed that the two-dimensional character of paper is no essential of computation. I assume then that the computation is carried out on one-dimensional paper, i.e. on a tape divided into squares. I shall also suppose that the number of symbols which may be printed is finite. If we were to allow an infinity of symbols, then there would be symbols differing to an arbitrarily small extent. The effect of this restriction of the number of symbols is not very serious. It is always possible to use sequences of symbols in the place of single symbols. Thus an Arabic numeral such as 17 or 999999999999999 is normally treated as a single symbol. Similarly in any European language words are treated as single symbols (Chinese, however, attempts to have an enumerable infinity of symbols). The differences from our point of view between the single and compound symbols is that the compound symbols, if they are too lengthy, cannot be observed at one glance. This is in accordance with experience. We cannot tell at a glance whether 9999999999999999 and 999999999999999 are the same. \cite{turing:machine}
\end{quote}

Turing begins with analyzing the way a human being normally calculates, that is by writing symbols on paper, and seems like he tries to analyze the essential aspects of it. Note that he does not analyze what happens in the human mind, he never refers to the human intellect. He tries to simulate the mechanical actions of the human being that calculates without taking into account the cognitive processes that take place during the calculation. We already see in the first lines of his analysis that Turing takes into account a one-dimensional tape as support for the calculations and not bi-dimensional one. In so doing, he eliminates the facilitations that the positional notation brings to the calculation process from a human being. Those facilitations can be seen as mainly cognitive as they do not add ``power'' to the calculation, the amount of functions that can be calculated in one-dimensional tape are the same as the amount of functions that can be calculated in a bi-dimensional paper (even though the mechanical steps made to calculate the same function may differ depending on the support that is used, but that does not concern us right now). From this point onward seems like Turing is slowly taking out of the model the human being from the model.

After that, he proceeds with an analysis of the set of symbols used during the calculation process, the alphabet of the machine if we prefer. That set must be a finite one and for a good reason. If we allow an infinity of symbols to be used then, from some point onward, the symbols will become too similar to one-another and distinguishing one symbol from the other could be very difficult. It's better if the symbols can be easily identified. By combining these symbols one can obtain an infinity of other compound symbols. An example that explains this concept are numbers. We can represent an infinity of numbers  just by combining a fixed set of ``primitive numbers''. We usually use a set of 10 numbers to represent all numbers:
\[ \left\{0, 1, 2, 3, 4, 5, 6, 7, 8, 9\right\} \]
If this set becomes too big it may become difficult to distinguish one element from the other, for example $12121212121212$ and $1212121212121212$. A second analysis of the symbol may be necessary.

\begin{quote}
The behavior of the computer at any moment is determined by the symbols which he is observing. and his "state of mind" at that moment. We may suppose that there is a bound $B$ to the number of symbols or squares which the computer can observe at one moment. If he wishes to observe more, he must use successive observations. We will also suppose that the number of states of mind which need be taken into account is finite. The reasons for this are of the same character as those which restrict the number of symbols. If we admitted an infinity of states of mind, some of them will be ``arbitrarily close'' and will be confused. Again, the restriction is not one which seriously affects computation, since the use of more complicated states of mind can be avoided by writing more symbols on the tape. \cite{turing:machine}
\end{quote}

The above-mentioned considerations can be also applied to the set of ``states of mind'' of the machine. Also, it's supposed that the number of observable symbols at a time is a finite one. If one desires to observe more than that quantity then consecutive observations are needed.

We can notice here a further attempt to simplify the model. Everything is described in terms of finite ``symbols''\footnote{Which may or may not have a particular meaning, the only concern is for them to be easily distinguishable from one-another.} and finite ``states of mind''. These restrictions, as Turing pointed out, do not reduce the power of the machine.

\begin{quote}
Let us imagine the operations performed by the computer to be split up into "simple operations" which are so elementary that it is not easy to imagine them further divided. \cite{turing:machine}
\end{quote}

At this point it becomes explicit Turing's attempt to make the operations the most simple ones possible.

\begin{quote}
Every such operation consists of some change of the physical system consisting of the computer and his tape. We know the state of the system if we know the sequence of symbols on the tape, which of these are observed by the computer (possibly with a special order), and the state of mind of the computer. We may suppose that in a simple operation not more than one symbol is altered. Any other changes can be set up into simple changes of this kind. The situation in regard to the squares whose symbols may be altered in this way is the same as in regard to the observed squares. We may, therefore, without loss of generality, assume that the squares whose symbols are changed are always ``observed'' squares. \cite{turing:machine}
\end{quote}

As indicated in the quote from above, and as easily verifiable, the state of a TM is known once the following are known:
\begin{enumerate}
\item The sequence of symbols written on the tape.
\item The observed symbol.
\item The current state of mind.
\end{enumerate}
One can notice that the machine is working more and more in a mechanical way. Each operation can be composed from simple alterations of the tape. The tape, as one can easily notice, is the sole component of the machine the content of which is mutable after the calculation has begun. The set of symbols and states of mind are not mutable.

\begin{quote}
Besides these changes of symbols, the simple operations must include changes of distribution of observed squares. The new observed squares must be immediately recognizable by the computer. I think it is reasonable to suppose that they can only be squares whose distance from the closest of the immediately previously observed squares does not exceed a certain fixed amount. Let us say that each of the new observed squares is within L squares of an immediately previously observed square. In connection with ‘immediate recognizability’, it may be thought that there are other kinds of square which are immediately recognizable. In particular, squares marked by special symbols might be taken as immediately recognizable. Now if these squares are marked only by single symbols there can be only a finite number of them, and we should not upset our theory by adjoining these marked squares to the observed squares. If, on the other hand, they are marked by a sequence of symbols, we cannot regard the process of recognition as a simple process. This is a fundamental point and should be illustrated. In most mathematical papers the equations and theorems are numbered. Normally the numbers do not go beyond (say) 1000. It is, therefore, possible to recognize a theorem at a glance by its number. But if the paper was very long, we might reach Theorem 157767733443477; then, farther on in the paper, we might find ‘... hence (applying Theorem 157767733443477) we have...’. In order to make sure which was the relevant theorem we should have to compare the two numbers figure by figure, possibly ticking the figures off in pencil to make sure of their not being counted twice. If in spite of this it is still thought that there are other ‘immediately recognizable’ squares, it does not upset my contention so long as these squares can be found by some process of which my type of machine is capable. [ \dots ] \\
The simple operations must therefore include:

(a) Changes of the symbol on one of the observed squares.

(b) Changes of one of the squares observed to another square within L squares of one of the previously observed squares.\\

It may be that some of these changes necessarily involve a change of state of mind. The most general single operation must therefore be taken to be one of the following:

A. A possible change (a) of symbol together with a possible change of state of mind.

B. A possible change (b) of observed squares, together with a possible change of state of mind. \cite{turing:machine}
\end{quote}

In addition to the manipulation of the symbols contained in the tape the ``computer''\footnote{The computer here intended is the human being that is calculating.} must also move the tape to read the next symbol. To facilitate this operation it is assumed that the next square of the tape is distant not more than $L$ units.\footnote{$L$ is intended as a finite quantity.}

Considering the assumptions made so far, we come to the conclusion that the mecanichal operations that a ``computer'' can make are of two types:
\begin{itemize}
\item A possible reading/writting of a symbol from/in the observed square of the tape with a possible change of the state of mind.
\item A possible change of the observed square with a possible change of the state of mind.
\end{itemize}

If we could draw an analogy, the calculation model proposed from Turing is, so far, similar to a musician playing a musical instrument. In order to render the explanation more comprehensible let's take for example a musician playing a piano in front of an audience. The written notes and the sound of the notes constitute the alphabet of the machine. The notes written in the copybook and given to the musician before the performance constitute the input of the calculation procedure. The pressing of the keys and the consequent sound produced by the piano constitute the result of the calculation procedure and the resulting melody constitutes the result of the calculation procedure. The creation of the melody is something not related to the instrument, the instrument is needed only in transforming the notes given as input in a melody. The possible physical operations in this case are:
\begin{itemize}
\item The reading of the notes given as input.
\item Press the key associated to a given musical note and eventually read the next symbol.\\
\end{itemize}

\begin{quote}
The operation actually performed is determined, [ \dots ], by the state of mind of the computer and the observed symbols. In particular, they determine the state of mind of the computer after the operation is carried out. We may now construct a machine to do the work of this computer. To each state of mind of the computer corresponds an "m-configuration" of the machine. The machine scans $B$ squares corresponding to the $B$ squares observed by the computer. In any move the machine can change a symbol on a scanned square or can change anyone of the scanned squares to another square distant not more than $L$ squares from one of the other scanned squares. The move which is done, and the succeeding configuration, are determined by the scanned symbol and the m-configuration. The machines just described do not differ very essentially from computing machines as defined in \S2, and corresponding to any machine of this type a computing machine can be constructed to compute the same sequence, that is to say the sequence computed by the computer. \cite{turing:machine}
\end{quote}

Taking into account all the assumptions afforementioned, we notice that the calculation process can easily be mechanized. As a matter of fact, each instruction of the machine can be represented as a tuple.\footnote{The choice made in this work is to represent each instruction as a 5-tuple, however, anyone who knows the TM, even if just a little, knows that they can be represented by different tuples.}

As has been suggested above, Turing has analyzed and simplied to the maximum the mechanical operations a human being executes when calculating, without taking into account its intelligence.\footnote{Which, so far, is a characteristic of the human being.} As a result of this operation he got some elementary operations which, combined with one-another, can produce more complex operations. Obviously, if the calculation process\footnote{Made-up from a combination of such elementary operations.} of a number ends in a finite amount of time and returns the desired result then the corresponding function is calculable, otherwise it may not be so.

By analysing the second point:
\begin{quote}
II. [Type (b)]. 
If the notation of the Hilbert functional calculus is modified so as to be systematic, and so as to involve only a finite number of symbols, it becomes possible to construct an automatic machine K which will find all the provable formulae of the calculus. \dots

\dots When sufficiently many figures of P have been calculated, an essentially new method is necessary in order to obtain more figures. \cite{turing:machine}
\end{quote}
it becomes clearer, from a mathematical point of view, what is calculable and what is not.

In the third point, Turing introduces another mean which substitutes the ``state of mind'':
\begin{quote}
III. This may be regarded as a modification of I or as a corollary of II.
We suppose, as in I, that the computation is carried out on a tape; but we avoid introducing the ``state of mind'' by considering a more physical and definite counterpart of it. It is always possible for the computer to break off from his work, to go away and forget all about it, and later to come back and go on with it. If he does this he must leave a note of instructions (written in some standard form) explaining how the work is to be continued. This note is the counterpart of the ``state of mind''. We will suppose that the computer works by such a desultory manner that he never does more than one step at a sitting. The note of instructions must enable him to carry out one step and write the next note. Thus the state of progress of the computation at any stage is completely determined by the note of instructions and the symbols on the tape. That is, the state of the system may be described by a single expression (sequence of symbols), consisting of the symbols on the tape followed by A (which we suppose not to appear elsewhere) and then by the note of instructions. This expression may be called the ``state formula''. We know that the state formula at any given stage is determined by the state formula before the last step was made, and we assume that the relation of these two formulae is expressible in the functional calculus. In other words we assume that there is an axiom U which expresses the rules governing the behavior of the computer, in terms of the relation of the state formula at any stage to the state formula at the proceeding stage. If this is so, we can construct a machine to write down the successive state formulae, and hence to compute the required number. \cite{turing:machine}\\
\end{quote}
In so doing, the power of the machine is in no way altered. This is just a more practical way of doing what the machinery\footnote{The mechanism by which the TM is composed, human beings calculating, in this case.} already did by using a note of instructions instead of the ``state of mind''.\footnote{Seems almost like the multitasking principle.}

The Church-Turing thesis can be summarized in:
\begin{quote}
For every effectively calculable function exists a TM that calculates it. The set of effectively calculable functions is identifiable with the set of recursive functions.\\
\end{quote}

The functions that are not calculable are the functions for which there exists no TM that calculates them. An example of this kind of function is the Halt function.

According to Copeland:
\begin{quote}
A method, or procedure, M, for achieving some desired result is called ‘effective’ or ‘mechanical’ just in case
\begin{enumerate}
\item M is set out in terms of a finite number of exact instructions (each instruction being expressed by means of a finite number of symbols);
\item M will, if carried out without error, produce the desired result in a finite number of steps;
\item M can (in practice or in principle) be carried out by a human being unaided by any machinery save paper and pencil;
\item M demands no insight or ingenuity on the part of the human being carrying it out. \cite{sep:ctt}
\end{enumerate} 
\end{quote}

And it seems like Turing thought so too, judging from what he says in \cite{turing:intelligent}:
\begin{quote}
A man provided with paper, pencil, and rubber, and subject to strict discipline, is in effect a universal machine.
\end{quote}

It's important to notice that this thesis is universally accepted but can not be effectively demonstrated.

The interested reader is adviced to read \cite{sep:ctt}, \cite{sieg:computability}, \cite{turing:machine}, \cite{sep:tm} and \cite{syropoulos:hypercomputation} for more information regarding the topic.


\section{Some models of Turing Machines.}

There are different models of TM's, in addition to the model presented by Turing himself. The purpose of this section is to make a brief introduction to some of these models. The power of the models presented in this section has been proven to be the same of the classical TM. Some other models will be mentioned later in the chapter regarding hypercomputation.

\begin{enumerate}
\item \textbf{Multitape TM.}

	The \emph{Multitape TM} is a TM with multiple tapes (as the name implies). As such, it differs from the classical TM by its transition function. The transition function for a TM with $n$ tapes becomes:
	\[ \delta: S \times A^n \to S \times A^n \times \left\{l, n, r \right\} \]

\item \textbf{TMs with a bidimensional tape.}

	The \emph{TM with a bidimensional tape} is a classical TM which has a tape which has infinite rows and infinite columns. The scanning head can move also \textbf{u}p or \textbf{d}own, this means that the transition function of these machines is:
	\[ \delta: S \times A \to S \times A \times \left\{l, n, r, u, d \right\} 	\]

	This type of TM allows an easy simulation of Multitape TMs and graphical elaborations. Further variants allow the use of 3D tapes.

\item \textbf{Nondeterministic TM.}

	The \emph{Nondeterministic TM} is a TM that can have multiple transitions for each combination of state of mind and input. Based on this model, the \emph{Probabilistic TM} model has been proposed.

\item \textbf{Simplified TM.}

	The \emph{Simplified TM} is a TM simplified in one of the following three aspects:
	\begin{enumerate}
	\item The tape.
	\item The alphabet.
	\item The states of mind.
	\end{enumerate}
	
	A TM can have an unlimited tape only by one side without losing any of its calculating power. 

	A TM can have an alphabet made by only two symbols without losing any of its calculating power.

	A TM can have a set of states of mind made by only two states without losing any of its calculating power.

	Note that each of the above simplifications can not be made at the same time with another simplification of the above.
\end{enumerate}

There are also several other models of TMs which are of interest, like:
\begin{enumerate}
\item Oracle Machines.
\item Infinite Time TMs.
\item Trial-And-Error Machines (TAE).\\
\end{enumerate}
a brief description of which will be given in the chapter regarding hypercomputation.

\section{Phisical interpretation of the TM steps.} \label{tm:steps}

The TM is, basically, an idealistic machine. It serves as a calculation model but it can not be built.\footnote{At least, not as it is described by Turing.} In the description of the TM, two limitations are not considered:
\begin{itemize}
\item \textbf{Time}: there is no upper bound to the computation time needed by a TM (even though the time at our disposal is limited).
\item \textbf{Space}: the tape of the TM is considered to be infinite/unbounded.\\
\end{itemize}

The number of steps executable by a TM is considered, as with time, to be finite even though it is seen as unbuonded. Regarding space, if we consider as true the results of the WMAP project of NASA, the observable universe is finite, even though it can grow limitlessly. During the computation of a TM one can create the illusion of an infinite tape just by adding more tape when it is needed, but the quantity of the tape available will always be finite. This is possible for the mere reason that the tape needed for the completion of the computation (assuming that the computation will end at some point) will always be a finite quantity.

If we want to consider time as a limited quantity (we put an upper limit to the computation time), automatically we put a limit to the number of steps a TM can execute. In this case we are interested in executing the greatest possible number of steps in a time unit. To do so one can operate in two ways:
\begin{itemize}
\item Try to use the minimal number of steps by using already existing or by creating new asymptotically optimal algorithms,
\item Increase the number of steps executable in a single time unit.
\end{itemize}

By choosing the second way one must understand that to achieve the desired result means to increase the speed of computation. As we can perceive by intuition, such speed can not be increased as much as one might desire because of physical reasons. But what are these reasons exactly?

A natural limit regarding speed, as the theory of relativity tells us, is the speed of light. As a matter of fact, as such theory tells us, $c = 299.792.458 \; m/s$ (in vacuum conditions) is the maximum speed information and matter can travel with in the universe. Obviously, while it is passing any material the speed of light is smaller than the previously indicated quantity.\footnote{Such quantity depends on the refractive index of the material itself.} The existence of the limit of the speed of light (a finite quantity) puts a limit to the maximum speed with which a computer can operate. Information, as a matter of fact, must travel between physical parts (circuits in today's computers). This claim is verified by the theory of relativity itself. In fact, the energy $E$ of an object with mass $m > 0$ and speed $v$ is given by the formula:
\[ E = \gamma m c^2 \]
where $\gamma$ is Lorentz's coefficient and has a value of:
\[ \gamma = \left( 1 - \frac{v^2}{c^2} \right)^{\frac{1}{2}} \]
As one can verify from the given formula, $\gamma = 1$ for $v = 0$ (giving life to the famous formula $E = m c^2$) and tends to infinity for $v$ that tends to $c$. This means that to make objects with positive mass travel at the speed of light one must use an infinite quantity of energy.

\begin{quote}
The speed of light is the upper limit for the speeds of objects with positive rest mass. \cite{fowler:relativity}
\end{quote}

This however is not all there is to say on the subject. As we all know, in the TM we also have interactions between physical parts. These interactions have also some thermodynamic effects. As Mundici explains in \cite{mundici:limitations}\footnote{See also \cite{mundici:machines}.}, the power consumption of a TM $T$ grows not less than the square of its frequency, or rather:
\[ f^2 \leq 2 \pi \frac{W}{h} \]
where
\begin{itemize}
\item $f$ is the frequency of $T$ (the number of steps executed in a time unit)
\item $W$ is the power used by $T$, measured in watt (power $=$ energy per second)
\item $h$ is Planck's constant\\
\end{itemize}
By the Heisenberg inequality, we have that the energy uncertainty must be:
\[ \Delta E \geq h \left( 2 \pi \Delta t \right)^{-1} \]
where:
\begin{itemize}
\item $\Delta E$ is the energy uncertainty
\item $\Delta t$ is the time needed for the execution of a single step.\\
\end{itemize}

The energy used for the step must be greater than $\Delta E$.

In their work, Mundici and Sieg conclude that the required volume $(V)$ to contain $z$ symbols must be:
\[
V \geq \frac{4}{3} z \pi a^3 m^3
\]
where:
\begin{itemize}
\item $a$ is the hydrogen's radius.\\
\end{itemize}

From this result we can deduce that the distance $d$ from two symbols must be:
\[
d = 2 r \geq 2 a z \frac{1}{3}
\]

Considering the speed with which the signals can travel (less than the speed of light) we have that the frequency $f$ is:
\[
f \leq c \left( 2 a z \frac{1}{3} \right)^{-1} \mathrm{steps \; per \; second}
\]

from which we have:
\begin{equation} \label{fz}
f z \frac{1}{3} \leq \frac{\left( \frac{a}{c} \right)^{-1}}{2} = \left( \frac{1}{2} \right) \times 5.655 \times 1018
\end{equation}

From (\ref{fz}) we can deduce that the maximum frequency of a machine is a finite quantity and it decreases while the number of symbols used by the machine increases.

We have now found some physical limits (mechanical ones) imposed on the machine. We have that the speed of signal propagation and the speed with which the mechanical parts of the machine can move have an upper bound, the speed of light $c = 299.792.458 \; m/s$. We also have that the energy consumption for a single step of the computation is lower bounded, the frequency of the machine is upper bounded and the volume in which a symbol can be contained is lower bounded.

Now that we know these limits we can proceed and talk about hypercomputation.

\chapter{Hypercomputation.}

In this section i will try to make a brief introduction to the notion of hypercomputation, continuing with a very brief description of the TM seen from the hypercomputation's point-of-view and then with a presentation of some models of hypercomputers. While introducing these models, an effort will be made analyzing the practical realizability of machines based on these models and then the section will come to an end with the introduction of some critics made to the notion of hypercomputability itself.

\section{An introduction to Hypercomputation.} \label{hypercomputation:intro}

Hypercomputation is a relatively new branch of calculability. It rises from the idea that the Church-Turing Thesis (CTT from now on) may not be true. The CTT specifies that for each effectively computable function exists a TM that computes it. The problem is that one can not demonstrate the truthfullness of the CTT.

In 1960 Scarpellini wrote, in a paper published by a german magazine in 1963, that the existence of non recursive precesses (processes which are not computable by a TM) may be possible in nature. He wrote:
\begin{quote}
One may ask whether it is possible to construct an analogue-computer which is in a position to generate functions $f(x)$ for which the predicate $\int f(x) \cos nx dx > 0$ is not decidable [by Turing machine] while the machine itself decides by direct measurement whether $\int  f(x) \cos nx dx$ is greater than zero or not.
\end{quote}

Scarpellini made it clear that:
\begin{quote}
Such a machine is naturally only of theoretical interest, since faultless measuring is assumed, which requires the (absolute) validity of classical electrodynamics and probably such technical possibilities as the existence of infinitely thin perfectly conducting wires. All the same, the (theoretical) construction of such a machine would illustrate the possibility of non-recursive natural processes.
\end{quote}
and then he proceeded with a consideration regarding the brain and hypercomputation:
\begin{quote}
It does not seem unreasonable to suggest that the brain may rely on analogue processes for certain types of computation and decision-making. Possible candidates which may give rise to such processes are the axons of nerve cells.
 
\dots

It is conceivable that the mathematics of a collection of axons may lead to undecidable propositions like those discussed in my paper.
\end{quote}

Hypercomputers compute functions which are not computable by TMs. This statement implies the fact that a hypercomputer may be able to compute the $halting$ function of a TM. This statement may seem to clash against the Turing's demonstration of the non-computability of such function but, as we will see, this is not the case. Turing's demonstration is quite interesting because it is applicable to entire \emph{classes} of machines. Basically, a TM can not compute the $halting$ function of any machine which is equivalent to a TM, but this does not imply that a machine which is not equivalent to a TM - of a different class from that of a TM - can not compute the $halting$ function of a TM. There is no logical contradiction in a machine of a certain class computing the $halting$ function of a machine belonging to a different class. A hypercomputer can not compute the $halting$ function for machines belonging to its own class.\footnote{The demonstration of this statement has been given by T. Ord and T. D. Kieu in \cite{kieu:diagonal}}

The first model of a hypercomputer is considered the machine introduced by Turing in his paper \cite{turing:oracle}, the \emph{Oracle Machine}.  This machine will be seen later in this work. After this machine there have been introduced several other models of hypercomputers.

Some hypercomputers are able to compute the so-called \emph{Supertasks}.\footnote{With the term \emph{Supertask} is indicated a task made by a numerable infinity of operations which are completed in a finite amount of time.} There are - as one might guess - several models of hypercomputers and in this work i will try to introduce and analize some of them and point out some of their limits.

It is important to clarify why the models of hypercomputers introduced in this work are more powerful than a TM. Saying it simply, a hypercomputer is similar to a TM with less limitations. If we notice, the limitations imposed on the original model of a TM - where the computer is a human being which computes - are:
\begin{itemize}
\item The computer does not use any form of intellect.
\item The computer can use only a pen and paper.
\item The computer follows a set of determined rules during the computation.
\item The computer has the data regarding the computation available to him written in some determined form in the paper.
\item The computer has the input data available before the computation starts.
\item The computer can not accept any input data after the computation has started.
\item If the computation has a result then it is unambiguous and it is obtained in a finite amount of time.\\
\end{itemize}

The simulation of a human being subject to the before-mentioned limitations - and of course the limitations given by the laws of physics - by a mechanical machine is what we know as a \emph{computer}. If we loose some limitations we can build a model of a hypercomputer. Each model of hypercomputation introduced in this work is basically a TM with less limitations (except one case).

\section{The TM seen from Hypercomputation.}

As it was pointed out before, a TM is just a calculating machine made by an infinite tape, a device to read from and write into the tape, a finite set of symbols - its alphabet - and a finite set of states of mind. The CTT says that this machine computes all the naturally computable functions, these functions being the functions computable by a human being endowed with a pen and paper, precise computing rules and not allowed to use its characteristic intelligence. This machine executes a finite set of operations:
\begin{enumerate}
\item Move the read/write device to the left or to the right.
\item Read from a cell of the tape.
\item Write to a cell of the tape.
\item Change the state of mind of the machine.\\
\end{enumerate}

About the way these operations are executed Turing, in \cite{turing:machine}, does not say anything.\footnote{Are they the result of the computation of a number or are they something else?} These operations are made available by some black boxes.\footnote{The composition and the functioning of these boxes is not necessary for the description of the machine itself.} In the year 1939 Turing publishes a paper in which another machine is described, what would have been later known as the first hypercomputer, the \emph{Oracle-machine} (O-machine).

\section{Some models of Hypercomputers.}

There exist several models of hypercomputational machines. The purpose of this section is not to make an exhaustive list of these models but to introduce the most relevant models and to analyze the main aspects regarding the realization of these machines, if possible. It must be clear that the last word regarding the realization of these machines is to be said by physics. Some of these models, according to modern physics, can not be built because they need some resources which are not available. Some of these models require an effectively infinite tape (it is not enough to have an illimited tape), some models require faultless measuring and for other models is required to work in a different time-space (more details will come afterwards). It seems though that there has been built a machine which is based on a hypercomputational model and this model will be discussed in the third chapter.

\subsection{O-machine.}

The \emph{O-machines} have been introduced by Turing in his paper \cite{turing:oracle}. The O-machines are TMs equiped with a black box capable of solving problems which are not solvable by a classical TM. These black boxes are, basically, machines that can hypercompute. This is how Turing introduced the O-machines:
\begin{quote}
Let us suppose that we are supplied with some unspecified means of solving number-theoretic problems; a kind of oracle as it were. We shall not go any further into the nature of this oracle apart from saying that it cannot be a machine. With the help of the oracle we could form a new kind of machine (call them o-machines), having as one of its fundamental processes that of solving a given number-theoretic problem.
\end{quote}

With the term "number-theoretic problem" are indicated the problems which can be formulated in arithmetical terms. We can suppose that, with this term, Turing meant to indicate problems similar to what he introduced in the 1936, to tell, given a TM, if it prints a finite quantity of binary digits or not.\footnote{If this was not the case then there would be no need for the existence of the Oracle because the problem would be solvable with a simple TM.}

Turing's introduction of the O-machines is, from a certain point of view, a little bit confusing. In it, Turing specifies that the oracle can not be a machine and, on the other hand, he describes them as "\emph{means}" and talks about the O-machines as "\emph{a new kind of machine}". Truth is, it may be difficult to consider the basic components of machines as machines themselves,\footnote{We do not consider a tape, in the case of a TM, as a machine.} but it is also difficult finding a "means" which is not a machine of some kind which is able to compute what is not computable by a TM. Maybe Turing meant that is could not be a TM, but the fact is that the oracle must be something that computes functions which are not computable by a TM, to say it otherwise: the oracle must compute more functions than a TM.

\subsection{TAE machine.}

The idea behind Trial-And-Error (TAE) machines was introduced in the 1965 by H. Putnam and M. Gold in their separate works \cite{jstor:tae} and \cite{jstor:limrec}. Basically, TAE machines are, as mentioned previously, TMs with less limitations, or more freedom if we want. This is easily understood by Putnam's introduction:
\begin{quote}
What happens if we modify the notion of a decision procedure by:

	\begin{enumerate}
	\item Allowing the procedure to "change its mind" any finite number of times (in terms of Turing Machines: we visualize the machine as being given an integer (or an n-tuple of integers) as input. The machine then "prints out" a finite sequence of "yesses" and "nos". The last "yes" or "no" is always to be the correct answer.); and
	\item We give up the requirement that it be possible to tell (effectively) if the computation has terminated? \\
	\end{enumerate}

I.e., if the machine has most recently printed "yes", then we know that the integer put in as input must be in the set unless the machine is going to change its mind; but we have no procedure for telling whether the machine will change its mind or not. \dots

If we always "posit" that the most recently generated answer is correct, we will make a finite number of mistakes, but we will eventually get the correct answer. (Note, however, that even if we have gotten to the correct answer (the end of the finite sequence) we are never sure that we have the correct answer.)
\end{quote}

Peter Kugel, as suggested in his article \cite{kugel:mind}, says that the human mind is a TAE machine (at least some parts of it). In fact, the way we think and act is sometimes similar to the way a TAE machine works, even though my personal opinion is not completely compatible with Kugel's in this case. The human mind also exhibits some characteristics of other models of hypercomputation later described in this work, like the \emph{Coupled Turing Machine (CTM)} and the \emph{Accelerated Turing Machine}. I agree with the fact that single events may be treated like a TAE machine does a single computation but let's recall the way these machines work. TAE machines receive the input data and compute the received input without accepting other input as the mind - and the CTM - does. Also, it seams that the "computations" - if we want to call them like this - executed by the mind are done faster each time one of them is repeated, which is a characteristic of ATMs.\footnote{Even though, as we will later see, the acceleration expected in this model is by far superior to the acceleration observed in the mind's computation.} In the section regarding the CTMs I will describe my opinion in more detail so that it can be more easily understandable. For more information regarding Kugel's model the reader should consult Kugel's work: \cite{kugel:mind}.

To better understand the way TAE machines work one should be familiar with the concept of \emph{Trial-And-Error procedure}. The \emph{Trial-And-Error} is an experimental method used for problem solving, repairing and knowledge acquisition. Like the name itself suggests, it is expected to try one possible solution after another until the problem is solved, keeping trace of the mistakes made previously.

This kind of approach is used for the resolution of simple problems or when none of the other approaches works. This does not mean that this approach is a brute force approach, on the contrary, most of the times there is a certain logic behind each step, until the reach of a successful result. Usually this method is put to use when one has little experience - or no experience at all - in the field the problem belongs to.

As described in \cite{cimbebasia:spiders}, it seems that some spiders use Trial-And-Error tactics to hunt for preys they have never seen before or when they are in unusual situations and they seem to remember the new tactics they have applied.

The \emph{Bogosort}, an extremely inefficient sorting algorithm, can be seen as a kind of Trial-And-Error algorithm, even though the original version of it does not keep trace of the already used combinations, violating in this way one of the principles of the Trial-And-Error procedures. The pseudo-code of this algorithm is:

\begin{verbatim}
while(not_sorted(array[])){
    scramble_elements(array[]);
}
\end{verbatim}

An effectively Trial-And-Error version of this algorithm, one that would keep trace of the already used combinations, would be more efficient than the original version. Such version would - in fact - guarantee that the procedure would end in a finite amount of time, something that the original version does not guarantee.

The Trial-And-Error is a method sometimes used to find new medicines. Sometimes two or more drugs are combined until a desired effect is obtained.

A simple application of this method was given in \cite{ashby:brain} (section 11/5). 
\begin{quote}
Suppose $N$ events each have a probability $p$ of success, and the probabilities are independent. An example would occur if $N$ wheels bore letters $A$ and $B$ on the rim, with $A$'s occupying the fraction $p$ of the circumference and $B$'s the remainder. All are spun and allowed to come to rest; those that stop at an $A$ count as successes. Let us compare three ways of compounding these minor successes to a Grand Success, which, we assume, occurs only when every wheel is stopped at an $A$.

\emph{Case 1}: All $N$ wheels are spun; if all show an $A$, Success is recorded and the trials ended; otherwise all are spun again, and so on till ' all $A$'s ' comes up at one spin.

\emph{Case 2}: The first wheel is spun; if it stops at an $A$ it is left there; otherwise it is spun again. When it eventually stops at an $A$ the second wheel is spun similarly; and so on down the line of $N$ wheels, one at a time, till all show $A$'s.

\emph{Case 3}: All $N$ wheels are spun; those that show an $A$ are left to continue showing it, and those that show a $B$ are spun again. When further $A$'s occur they also are left alone. So the number spun gets fewer and fewer, until all are at $A$'s.

\dots

\dots Suppose, for instance, that $p$ is $\frac{1}{2}$, that spins occur at one a second, and that $N$ is 1000. Then if $T_1$, $T_2$ and $T_3$ are the average times to reach Success in Cases 1, 2 and 3 respectively, 
\[ T_1 = 2^{1000} \; \mathrm{seconds,} \]
\[ T_2 = \frac{1000}{2} \; \mathrm{seconds,} \]
\[ T_3 = \mathrm{rather \; more \; than \;} \frac{1}{2} \mathrm{\; second.} \]
\end{quote}

Notice that no kind of intelligence is used in the described decision procedures. All 3 procedures are purely mechanical.

Ashby, noticing the different solutions based on a Trial-And-Error strategy, suggested the existence of meta-levels of Trial-And-Error, a kind of TAE hierarchy if we can say so. This idea is extended to the point of having a recursive sequence of meta-levels. Based on these consideration, Ashby comes to the conclusion that human intelligence is based on TAE trials. In other words, Ashby suggests that there is are some primitive TAE trials upon which a certain number of meta-levels are built until we have what we now know as Intelligence. One comes to the conclusion that Intelligence is nothing more than a set of TAE trials.

Regarding TAE machines, we can say that these machines are able to execute decision procedures applied to recursively enumerable sets. As we know, these decision procedures, generally, can not be executed by a TM until we violate at least one of the conditions introduced in the section \ref{hypercomputation:intro}.

As we know, recursively enumerable sets are supersets of the recursive sets. In other words, the recursively enumerable sets are the sets that have a partially recursive characteristic function, one of the following kind:
\[
f(x) = \left\{\begin{matrix}
	1 & if \; x \in A \\
	\mathrm{indefinite} & otherwise
\end{matrix}\right.  
\]

The reader interested in more detailed informations regarding recursively enumerable sets can read \cite{ausiello:linguaggi} and \cite{brainerd:computation}.

In his work, Putnam defines also the concept of \emph{Trial-And-Error predicate}:

\begin{quote}
$P$ is a trial and error predicate if and only if there is a g.r. (general recursive) function $f$ such that (for every $x_1, x_2, \dots, x_n$)
\[ P(x_1, x_2, \dots, x_n) \equiv \lim_{y \to \infty} f(x_1, x_2, \dots, x_n, y) = 1 \]

\[ \bar{P}(x_1, x_2, \dots, x_n) \equiv \lim_{y \to \infty} f(x_1, x_2, \dots, x_n, y) = 0 \]

where

\[ \lim_{y \to \infty} f(x_1, x_2, \dots, x_n, y) = k \overset{\underset{\mathrm{def}}{}}{=} \exists y \forall z (z \geq y \Rightarrow f(x_1, x_2, \dots, x_n, z) = k) \]
\end{quote}

Putnam introduced also the concept of a \emph{k-trial predicate} in the following way:

\begin{quote}
Call a predicate $P$ a \emph{k-trial predicate} if there is a g.r. function $f$ and a fixed integer $k$ such that (for all $x_1, x_2, \dots, x_n$)
\[ P(x_1, x_2, \dots, x_n) \equiv \lim_{y \to \infty} f(x_1, x_2, \dots, x_n, y) = 1 \]
\end{quote}

Based on Putnam's results, one can say that a predicate is a Trial-And-Error one if and only if a recursive function which gives a result, in a finite number of steps, which will not change anymore, exists.

In the same paper, Putnam gave proof of the following theorem:

\textbf{Theorem:} $P$ is a Trial-And-Error predicate if and only if: \[ P \in \Pi_2 \cap \Sigma_2 \]

Notice that the functions computable by a TM are the recursive ones ($\Delta_1$). A TAE predicate is an application of the limit to a $\Delta_1$ predicate, in other words, a $\Delta_2$ predicate.

A $\Sigma_n^0$ predicate is a predicate which has the following form:
\[ \exists n_1 \dots \exists n_k \psi \mathrm{\; where \;} \psi \mathrm{\; is \; a \;} \Pi_{n-1}^0 \mathrm{\; predicate.} \]

A $\Pi_n^0$ predicate is a predicate which has the following form:
\[ \forall n_1 \dots \forall n_k \psi \mathrm{\; where \;} \psi \mathrm{\; is \; a \;} \Sigma_{n-1}^0 \mathrm{\; predicate.} \]

A $\Sigma_1^0$ predicate is a predicate which has the following form:
\[ \exists n_1 \dots \exists n_k \psi \]
where $\psi$ contains quantifiers of the form:
\[ \forall n < t \mathrm{\; and \;} \exists n < t \]
The sets:
\[ A \in \Sigma_1^0 \]
are the recursively enumerable sets. An example of such set is:
\[ A = \left\{ x | \exists i \in \mathbb{N}, \varphi_x(i) \; ends \; in \; less \; than \; i \; steps \right\} \]

The negation of a $\Sigma_1^0$ predicate is a $\Pi_1^0$ predicate. Proceeding with the previous example, we have:
\[ \bar{A} = \left\{ x | \not{\exists} i \in \mathbb{N}, \varphi_x(i) \; ends \; in \; less \; than \; i \; steps \right\} \]
\[ \bar{A} = \left \{ x| \forall i, \varphi_x(i) \; does \; not \; end \; in \; less \; than \; i \; steps \right \} \]

A $\Delta_n^0$ predicate is both a $\Sigma_n^0$ and a $\Pi_n^0$ predicate, more precisely:
\[ \Delta_n^0 = \Sigma_n^0 \cap \Pi_n^0 \]

The sets $T$ for which exists a TM that calculates the function:
\begin{equation} 
f(x) = \left\{ \begin{matrix} 1 & if \; x \in T \\ 0 & otherwise \end{matrix} \right. 
\end{equation}
are $\Delta_1^0$ sets.

The set of numbers computable by a TAE machine, as one can easily see from the previously given data, is from a superior level in the arithmetic hierachy than the set of numbers computable by a TM.

These machines can be used to calculate the halting problem - as described by Kugel - and other equivalent problems such as the verification of the Goldbach's conjecture. The Goldbach's conjecture can be expressed in the following way:

\begin{quote}
Every even integer greater than 2 can be expressed as the sum of two prime integers.
\end{quote}

To solve this problem a TAE machine could:
\begin{enumerate}
\item The machine prints "yes" as a first answer.\footnote{The first even integer greater than 2 is 4 which can be written as 4 = 2 + 2.}
\item The machine analizes the next even integer.
\item If the analized number can be written as the summ of two prime integers the machine continues from step 2, otherwise the machine "changes its mind" and prints "no" as the next answer and eventually terminates its computation.\\
\end{enumerate}

Other authors have introduced other hypercomputational machines similar to the TAE machines.\footnote{Jaakko Hintikka and Arto Mutanen in \cite{hintikka:math}(chapter 9) introduced a machine very similar to a TAE machine. Mark Burgin, in \cite{burgin:srec}, introduces the Inductive Turing Machines. For more information regarding these machine consult \cite{hintikka:math} and \cite{burgin:srec}.}

\subsection{Accelerated Turing Machine.}

The \emph{Accelerated Turing Machines} (ATM from now on) have been inspired by the Zeno's paradox. To describe this paradox let's give the classic example of the race between Achilles and the tortoise.

If the tortoise and Achilles must reach a certain place $B$ starting from a point $A$ and the tortoise starts with an advantage of some meters, Achilles will not be able to reach the finish before the tortoise. To reach the finishing line before the tortoise Achilles must first reach the tortoise but in the meantime the tortoise has advanced in further away point. To surpass the tortoise, Achilles must reach the actual position of the tortoise but then, again, the tortoise in the meantime has reached a further away point. Going on with this kind of reasoning one can easily conclude that Achilles will never reach the tortoise as he must reach the tortoise's location an infiniti of times. The paradox can be better explained using a simple picture:\\ \mbox{} \\
\includegraphics[width=\textwidth]{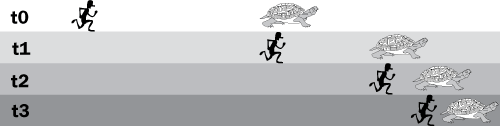} \\

The idea of accomplishing an infinite amount of actions\footnote{A numerable infinity.} in a finite amount of time is represented with the word \emph{supertask}. This word must not be confused with the word \emph{Hyper-task}  which represents the idea of accomplishing a non numerable infinity of actions in a finite amount of time. According to some philosophers, supertasks are logically impossible. The reason for this will be more clear after the introduction of the ATMs.

The ATMs are TMs that execute the first step of the calculation in $2^0 = 1$ unit of time, the second step in $2^{-1}$ units of time, the third step in $2^{-2}$ units of time, the fourth step in $2^{-3}$ units of time and so on. The time $t$ needed for the execution of $n$ steps of a computation is given by the (well known) formula:
\[ t = \sum_{i = 0}^{n} 2^{-i} \]
It is easy to verify that:
\[ \lim_{n \to \infty} t = 2 \]
This result says that this machine can execute an infinite amount of steps in 2 units of time.

Now, let's make the machine compute the function $x = x + 1$ for $n$ times where the initial value of $x$ is $0$ and $n \to \infty$. The first time this operation will be executed in $1$ minute.\footnote{We are taking a minute as the time unit.} The second time the operation will be executed in half a minute, $2^{-1} = \frac{1}{2}$. The third time  it will be executed in a quarter of a minute, $2^{-2} = \frac{1}{4}$, and so on. The $n$-th time the operation will be executed in $2^{-(n-1)}$ minutes. At the end of the second minute (the end of the execution), will $x$ be an even number or an odd one?

The main philosophical problem of the ATMs, but we can say that it is a problem which regards supertasks, is the fact that one knows the initial state of the machine - or state of the world, if we may say so - in the beginning of the computation but the final state is unknown. We'll go more in depth of this problem later when we'll discuss the power of these machines.

From a certain point of view, ATMs work as human beings - the more we do something the faster we become at doing it even though, at some point in time, one won't be able to become faster than what he already is.

I find it quite interesting to analyze these machines in two respects:
\begin{enumerate}
\item Physical realizability.
\item Computational power.
\end{enumerate}

From Physics we know that these machines are unrealizable. In section \ref{tm:steps} are described some physical limitations naturally imposed to each calculation step. As one may notice, these limitations are easily reachable by ATMs, even if one is willing to wait for a relatively long time. The difference between the number of calculation steps executed by this machine in $1$ minute and the number of calculation steps executed in $64$ minutes is just $6$. The difference between the number of calculation steps executed in $1$ second and the number of calculation steps executed in $2^{1000}$ seconds - a waiting time greater than the estimated age of the universe - is just $1000$. Assuming that the I/O head of the machine moves at a speed of $1 \frac{m}{s}$ for the first calculation step we would have that the speed necessary for the execution of the $29$-th calculation step would be greater than the speed of light, a rather low number of steps. Keeping in mind these considerations one can think of reducing the distance between the squares of the tape when the speed can not be further increased. Even in this case though we would have that after a few steps the distance would not be further reducible.

Let's suppose that we can overcome the limit of the speed of light.\footnote{Putz and Svozil have introduced in \cite{putz:light} the theoretical possibility of "pushing" a machine to compute with a greater speed than the speed of light by immersing it in a substance with a refractive index lower than one. As the authors say: \begin{quote} \dots at present such a possibility merely remains a theoretical speculation \end{quote}} In this case another problem arises: space. A function may need a finite - arbitrarily large - quantity of tape but, in the case of ATMs, it may also need an infinite quantity of tape. Let's consider the following programs, written in pseudo-code: \newline \newline 
\textbf{Program 1:}
\begin{verbatim}
begin
    i := 1;
    while(i > 0) do i := i;
end 
\end{verbatim} 

 \mbox{} \\ \mbox{} \\ 
\textbf{Program 2:} 
\begin{verbatim}
begin
    i := 1;
    while(i > 0) do i := i + 1;
end
\end{verbatim}

For the first program we have that the necessary quantity of tape for its computation is finite. This is not true for the second program. In fact, the second program needs an infinite quantity of tape, a quantity which is not containable in the observable universe.\footnote{Some considerations on this topic are given in the section \ref{tm:steps} of the present thesis.}

If we want an ATM with a finite quantity of tape - even if it may be arbitrarily large - we must give up on the idea of having it compute more than the classical TM. In \cite{calude:atm}, Calude and Staiger give proof that every TM that uses a finite quantity of tape - even if it operates as an ATM - can not compute a function which is not computable by a classical TM.

Let's talk now about the computational power of ATMs (without taking into account the problem of their physical realizability). As said previously, an ATM is able to execute supertasks and that brings forth some logical problems related to them. One of the most known is the problem of \emph{Thomson's lamp}. The problem introduced by Thomson\footnote{The British philosopher who introduced the term \emph{supertask}.} was to specify the state of a lamp that switched states - on/off - in a time-pattern similar to the execution of the computation steps in ATMs in the end of $2$ time units. The problem seems to be in the fact that if the lamp is \emph{off} at the end of the second time-unit it will switch to \emph{on} right after and vice versa. Benacerraf - in \cite{benacerraf:tasks} - noticed that this argument is invalid. In fact, we can see the operation of such a mechanism as a function specified in the range $\left[ 0, 2 \right)$. This function is not specified out of this range, that is the range $\left[ 2, \infty \right)$. The function may or may not be continuous for the value $2$.\cite{shagrir:atm}

There is another considerable problem with supertasks, its existence. Is there any way of executing a supertask? Theoretically it is possible to execute supertasks. As we will see later in this thesis, in this universe seems like there exist space-time structures that allow the execution of supertasks.

In \cite{shagrir:atm} the author argues about another topic: By making a machine able to execute supertasks does its computational power really increase?

\begin{quote}
Unlike the ``ordinary'' Turing machines, accelerating Turing machines can complete infinitely many steps within a finite span of time. But do they have more computational power? Do they solve, for example, the halting problem? I argue that they do not. \cite{shagrir:atm}
\end{quote}

In case the author, with the term \emph{halting problem}, intends \emph{halting problem for a classical TM}, I must disagree with what is said above. In \cite{kieu:diagonal}, as said before, is demonstrated that a machine is unable to solve its own halting problem or the halting problem of machines belonging to its class, but this does not prevent them from solving the halting problem of machines belonging to a different class. A way in which an ATM can solve the halting problem of a TM is described in T. Ord's work \cite{ord:hypercomputation}:

\begin{quote}
Consider an accelerated Turing machine, A, that was programmed to simulate an arbitrary Turing machine on arbitrary input. If the Turing machine halts on its input, A then changes the value of a specified square on its tape (say the first square) from a 0 to a 1. If the Turing machine does not halt, then A leaves the special square as 0. Either way, after 2 time units, the first square on A’s tape holds the value of the halting function for this Turing machine and its input.
\end{quote}

The problem, according to Shagrir, lies in the fact that an ATM does not have a state to indicate the end of the computation, as is the case with the Infinite Time Turing Machine. A simple version of this machine can be seen as an ATM which, at the end of the second unit of time enters a special state and then continues its work with the output of the previous 2 time-units computation as input for the next 2 time-units computation. According to Shagrir, a machine that has the same structure of a TM can compute the same functions of a TM even if it is able to execute supertasks.

\begin{quote}
I do not deny that this infinite time Turing machine computes the halting function. I also do not mind the name infinite time Turing machine. Rather, my point is this. If accelerating Turing machines have exactly the same computational structure as ordinary Turing machines, then they compute exactly the Turing machine computable functions. Performing supertasks enables the accelerating machines to complete infinitely many steps in a finite interval of time, but it does not enable to compute functions that the ordinary machines cannot compute. And if accelerating Turing machines differ from ordinary Turing machines in computational structure, as is the case with infinite time Turing machines, then they might have more computational power. But here too, the difference in computational power is not due to performing supertasks alone. Performing a supertask only ensures that the computation terminates in a finite real time, even if it requires infinitely many computation steps. The difference in computational power owes to the difference in computational end structure. Either way, no paradox emerges. If the accelerating Turing machine has the same computational structure as the ordinary machine, it does not compute the halting function. And if we extend the concept of the Turing machine, redefining the end structure, it should come as no surprise that the newly specified Turing machines compute functions, e.g., the halting function, that Turing's machines – the machines that Turing specified – fail to compute. \cite{shagrir:atm}
\end{quote}

I find that making machines able to enter a special state at the end of a fixed period of time means making them ``conscious'' about time or about 
the steps executed up until that moment.

Regarding the computational power of Infinite Time Turing Machines we have that:

\begin{quote}
There are no logical contradictions in the infinite time Turing machine metaphor and they have given rise to an interesting theory in which, for example, $P \neq NP$ and $\Pi_1^1$ sets are decidable by these devices. However, there is no suggestion at all of how such devices might be engineered or even conceived in a physical theory (nor is it necessary if these devices are considered in a logic context only) so these are ``machines'' in name only. \cite{potgieter:zeno}
\end{quote}

ATMs are able to decide predicates $P$ such that: $P \in \Pi_1^0 \; or \; P \in \Sigma_1^0$. These machines are able to compute the characteristic functions of the recursively enumerable sets.

\subsection{Real Computer.}

A \emph{Real Computer} (RC from now on) is a computer which process real numbers - $x \in \mathbb{R}$ - with infinite precision. This machine can be seen as an ideal analog computer. As one might guess, for the physical realization of this machine, means that let you work with numbers with infinite precision are needed.\footnote{Physics tells us that these means can not be built.}

There have been some proposals of considering the number of particles in the observable universe - a number estimated to be between $10^{79}$ and $10^{81}$ - as infinite. According to these proposals the number of decimal digits of a real number must have this length at most. We know that if we consider these proposals as reasonable we must consider the set of real numbers as a finite set, making it perfectly numerable. It would be such even if we consider it as an infinite set but made only with numbers which length would be arbitrary - even greater than $10^{81}$ - but finite. Such a number can very well be represented by a function which uses two natural numbers as parameters: 
\[ real(a, b) = a \times 10^{-b} \; \mathrm{where} \; a, b \in \mathbb{N} \]

Having such a function and considering the set of real numbers as an infinite set made only with numbers which length is arbitrary but finite would allow us to use the method used to counted fractions or, in general, each couple of natural numbers. Assuming that we have pairs of natural numbers that are supposed to represent real numbers in a matrix like the matrix used by Cantor to demonstrate the non-countability of real numbers, numbering its diagonals in the opposite direction of what is used for the demonstration of the non-countability of real numbers and assigning the number $0$ to the first pair, we would need only $3$ functions to enumerate the set of real numbers made as said before. These functions would be:

\begin{enumerate}
\item A function to estimate the number of the first couple in a diagonal $x$:
	\[ f(x) = \frac{x(x + 1)}{2} \; for \; x \in \mathbb{N} \]
	
\item A function to estimate the number that must be given to a couple of natural numbers $(x, y)$:
	\[ g(x,y) = \left\{\begin{matrix} 	0 & \; se \; x = 0\\ g(\frac{x}{10}, y - 1) & \; se \; x \; mod \; 10 = 0\\ f(x+y) + y & \; altrimenti \end{matrix}\right. \]

\item A function that estimates the values of the elements of a couple with a given representation number:
	\[ h(x) = \left ( \frac{\left \lfloor \frac{\sqrt{1 + 8x} - 1}{2} \right \rfloor \left \lfloor \frac{5 + \sqrt{1 + 8x}}{2} \right \rfloor}{2} - x, x - \frac{\left \lfloor \frac{\sqrt{1 + 8x} - 1}{2} \right \rfloor \left \lfloor \frac{1 + \sqrt{1 + 8x}}{2} \right \rfloor}{2} \right ) \] \\
\end{enumerate}

Assuming that we have the necessary means to measure a real number with arbitrary precision we would have all the necessary tools to manipulate the set of real numbers. Modern Physics tells us that these means are only ideal ones being that from a certain point onwards we would have some insurmountable physical problems with the measurements. The problem with the countability of the set of real numbers is the infinitely small numbers, those numbers that are represented by an infinite amount of decimal numbers and which are not periodic. If we limit the set of real numbers to those numbers that are not infinitely small we actually take out of the set those numbers that make it non-countable, but would it actually be useful to computability? I do not think so, after all, we are already able to count this set.

\subsection{Coupled Turing Machines.}

As we know, a TM does not accept any input once its computation has started. The idea behind the \emph{Coupled Turing Machines} (CTM from now on) is to have a TM that accepts input which may come even after the start of the computation. Theoretically, not all CTMs can be simulated by TMs. In the case of the computation of a real number, a CTM can operate on a digit of the number while it is receiving the next digit in input. This operation can not be executed by a TM because in this case the digits of the real number must first be generated - all of them - and then given as input to the TM so that it can operate on them.

As I personally see it, the CTM is like an O-machine where the environment with which the machine is connected plays the role of the oracle. The only feature that makes it different from an O-machine is the fact that the CTM's input can be a data stream.

As I mentioned earlier, the way this machine works is similar to the way our mind works. It is my intention to make a brief introduction to Kugel's model of the mind and to explain the reasons why I do not agree totally with this model.

According to Kugel, the mind is made of $4$ modules:
\begin{itemize}
\item \emph{Input Processor -} the module that gathers data from the surrounding environment and transforms them in data that are accessible and modifiable by the \emph{Central Processor}. For example, it can receive a visual signal from the retina and transform it in the message ``\emph{A tiger is near}'' and send it to the Central Processor for further processing.

\item \emph{Central Processor -} the module that receives the input data from the Input Processor and transforms it in a message that will be given as input to the \emph{Output Processor}. For example, the message ``\emph{A tiger is near}'' could be transformed in the message ``\emph{Run}''.

\item \emph{Program Selector -} the module that selects the program that has to be executed at a certain moment. For example, when a tiger is seen this module may choose to execute the \emph{ANIMAL IDENTIFICATION} program rather than the \emph{APPRECIATION OF THE BEAUTY} program.

\item \emph{Output Processor -} the module that receives as input the message sent by the Central Processor and transforms it into messages that can change the surrounding environment. For example, it can take the message ``\emph{Run}'' as input and transform it in messages that make it possible to control the different muscles that are necessary to actually run.\\
\end{itemize}

Although I fairly agree with the logical division in the $4$ modules listed above, I disagree with the fact that these modules are TAE machines. I would rather see them as CTMs. These modules are always working and accept input even when they are already computing, which is a characteristic of CTMs. I partially disagree with the data flow given in \cite{kugel:mind} (page 5) also. \\

\includegraphics[width=\textwidth]{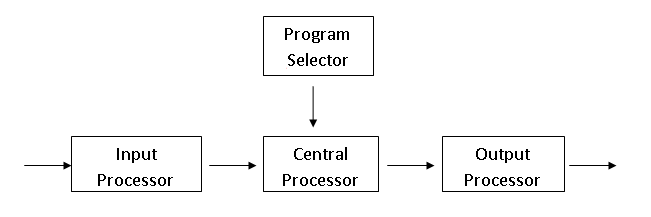} \\

In my opinion, the data flow given above is incomplete and should be like the following:\\

\includegraphics[width=\textwidth]{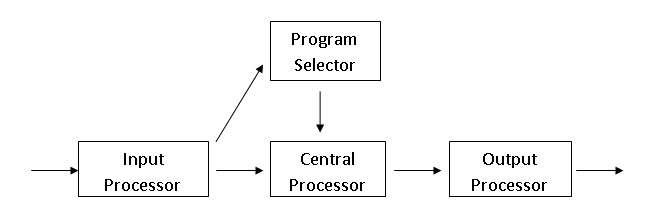} \\

The Input Processor gathers data from the surrounding environment and transforms them in data that are accessible and modifiable byte the Central Processor and Program Selector. The Central Processor receives the data from the Input Processor and Program Selector and then processes them and afterwards they are sent to the Output Processor. These modules - except for the Input Processor - can be seen also as TAE machines that accept input even while they are computing (a sort of Coupled TAE machine). The Input Processor's can be seen as a simple CTM as its job is to receive data continuously and make them accessible to other modules.

\subsection{Malament-Hogarth Machine.} \label{hypercomputation:mhm}

This model was created once it was observed that is not always necessary for a computer - a TM - to be under observation, the observer does not have the duty to continually keep an eye on the TM while it is computing. Theoretically it is possible to send a TM in a different space-time from the one the observer is in, in a space-time where it can execute an infinite amount of computational steps while for the observer has only passed a finite amount of time. One can put a signaling device in the TM which will send a signal to the observer when the computation has come to an end. In this way the TM can solve the halting problem: If the signal is sent before a certain (finite) time limit, that computation ends, if not, it does not end. 

I would say that for the production of this computing system there are a few considerations to make:

\begin{enumerate}
\item Finding - or creating, if possible - the environments with the necessary space-times.
\item Having the necessary means to move and comunicate from one space-time to the other.
\item Having the means that let you work safely.
\end{enumerate}

Quoting \cite{etesi:spacetime}:

\begin{quote}
The Kerr metric, which describes empty space-time around a rotating black hole, possesses these features: a computer can orbit the black hole indefinitely, while an observer falling into the black hole experiences an M-H event as they cross the inner event horizon. (This, however, neglects the effects of Black Hole Evaporation.)
\end{quote}

The issue becomes more complicated if one considers the effects of Black Hole Evaporation. 

For more information regarding this model see \cite{syropoulos:hypercomputation} and \cite{etesi:spacetime}.

\section{Critics to the notion of Hypercomputation.}

In his paper \cite{copeland:hypercomputation}, Copeland answers to several critics made to the notion of hypercomputation. The purpose of this section is to give a brief summary of some of these critics and the answers given. 

\begin{enumerate}
\item \begin{quote}
\emph{Any task that can be made completely precise can be programmed for the universal Turing machine. In other words, given enough memory and sufficient time, a standard digital computer can compute any rule-governed input-output function. That is what Turing and Church showed. Therefore the notion of hypercomputation is otiose.}
\end{quote}

\begin{quote}
\dots Turing and Church are sometimes said to have shown that a standard digital computer can, given enough memory and sufficient time, compute any rule-governed input-output function\dots In fact, they showed the opposite. There is nothing imprecise about the halting problem. The halting function is certainly rulegoverned.
\end{quote}

\item \begin{quote}
\emph{Turing showed in 1936 that every mechanical process can be carried out by the universal Turing machine. Therefore ‘hypercomputers’ are not machines of any sort — let alone computing machines.}
\end{quote}

Turing, in 1936, showed that a TM can execute all the operations that a human being could execute if he would be subject to the rules mentioned in \S1. He showed that the computable - mechanically, by a TM - numbers are all the numbers that are computable by a human being subject to the before-mentioned rules. 

\begin{quote}
This thesis carries no implication concerning the extent of what can be calculated by a machine, for among the machine’s repertoire of fundamental processes there may be those that a human rote-worker unaided by machinery cannot perform.\cite{copeland:hypercomputation}
\end{quote}

Personally, I agree only partially with the quote above. I think that the power of a machine is directly linked to the conditions it is working in. I do not think that - even though I am not totally sure about it - among the processes that are executable from a machine there may be processes that are not executable by a human being.

\item \begin{quote}
\emph{Hypercomputation seems to amount to the claim that there might be mechanical processes that are not mechanical!}
\end{quote}

\begin{quote}
True—so long as ‘mechanical’ means something different at the two occurrences. At the second occurrence, ‘mechanical’ has its technical sense: ‘not mechanical’ means ‘cannot be done by a human computer’. At the earlier occurrence, ‘mechanical process’ means simply ‘process that can be carried out by a machine’.
\end{quote}

\item \begin{quote}
\emph{Over the years, a number of alternative analyses have been given of the notion of a mechanical process. Apart from Turing’s analysis in terms of Turing machines, and Church’s analyses in terms of lambda-definability and recursiveness, there are analyses, e.g., in terms of register machines, Post’s canonical and normal systems, combinatory definability, Markov algorithms, and Gödel’s notion of reckonability. The striking thing is that these various analyses all turn out to be provably equivalent in extension. Because of the prima facie diversity of the various analyses, their equivalence is strong evidence that whatever can be done by a machine, mathematically speaking, can be done by the universal Turing machine.}
\end{quote}

The analyses under question are all analyses of the notion of an effective method. These analyses form a strong evidence that the Church-Turing Thesis is true but they do not say what the power of a machine may be if it operates under different conditions.\footnote{TAE machines are a perfect example. TAE machines are TM that operate under different conditions and they can calculate more functions than a TM.}

\item \begin{quote}
\emph{It seems that according to hypercomputationalists, every function is computable (or generatable by some machine). Each number-theoretic function is computable by a machine accessing an infinite tape on which are listed all the arguments of the function and the corresponding values. ETMs (Section 1.6) even permit an entire real number to be stored on a single square of the machine’s tape. And there is no reason to stop there -- additional fantasy brings additional computable functions. On the new way of speaking, ‘computable function’ means simply ‘function’. Hypercomputationalism comes down to this: the term ‘computable’ is redundant.}
\end{quote}

\begin{quote}
Hypercomputationalists believe that statements concerning computability are explicitly or implicitly indexed to a set of capacities and resources\dots When classicists say that some functions are absolutely uncomputable, what they mean is that some functions are not computable relative to the capacities and resources of a standard Turing machine. That particular index is of paramount interest when the topic is computation by effective procedures. In the wider study of computability, other indices are of importance. As the objection indicates, some indexed statements of computability are entirely trivial — for example, the statement that each number-theoretic function is computable relative to itself. This is not generally so, however. Mathematical theorems of the form ‘f is computable relative to r’ are often hard-won. Questions about which functions are computable relative to certain physical theories are seldom trivial. The question of which functions are computable relative to the theories that characterise the real world is of outstanding interest.
\end{quote}

\item \begin{quote}
\emph{One suggestion made by hypercomputationalists is that some form of quantum computer may be able to compute non Turing-machine-computable functions. However, the originator of the universal quantum computer, David Deutsch, states that this is not so\dots}
\end{quote}

\begin{quote}
A number of different quantum computational architectures have been proposed. Some are not hypercomputational, some are. In a paper in this collection, Kieu outlines a hypercomputational quantum computer that is able to solve Hilbert’s tenth problem.

Despite what Deutsch says, his universal quantum computer is able to compute non-recursive functions, since an entire non-recursive function can be encoded into one of the real-valued parameters figuring in the quantum-mechanical description of the machine (Solovay, personal communication)\dots
\end{quote}
\end{enumerate}

\chapter{Quantistic Computation.}

In the last part of the thesis we will talk about \emph{Quantum Computing} and its connection to Hypercomputation. We will start with a brief introduction to Quantum Computing so that we can give some basic information to better understand what will be introduced later, the \emph{Adiabatic Quantum Computing}.

Notice that the purpose of this chapter is not to give an exhaustive explanation of the concepts mentioned above; it is just to make a brief introduction to these concepts and give the necessary information regarding the objective of this thesis. For more information on these concepts one is advised to read \cite{williams:qc}, \cite{sep:quantum} and \cite{shankar:quantum} and the papers of Tien D. Kieu mentioned in the coming sections. 

\section{An introduction to Quantum Computing.}

Lately, in the scientific world, quantum computing is becoming more well--known, and for a good reason I would dare say. Gordon Moore -- one of the co--founders of Intel -- noticed in 1965 that the number of transistors per surface unit in a circuit was doubling every 18 months while the power increased. According to Moore, this trend would continue in the future and -- as we well know -- so it was. That affirmation became known as \emph{Moore's Law}. If this trend will continue then fewer atoms will be used to implement more bits, until one atom will be used to implement one bit. With the current trend this limit is estimated to be reached in the year 2020.

At that point, classical physics will not be sufficient to describe and handle the physical steps of a computation. At that point the use of Quantum Physics will be necessary. The laws of quantum physics are very different from the laws of classical physics. That which is normal for quantum physics is not normal at all for classical physics. For example, a quantum may be in more than one place at a time, or in more than one physical state at the same time. This behaviour is inconceivable for an object in the domain of classical physics. A quantum bit can have a value of \textbf{0} or \textbf{1} at the same time. Using quantum computing one can obtain a random number while in classical computation one can only obtain a pseudo--random one. The microscopic objects described by quantum mechanics behave sometimes like particles and sometimes like waves.

The topics of most interest for us right now are: \emph{What can a Quantum Computer compute? Can a Quantum Computer compute more than a TM?}

According to David Deutsch -- the creator of the Universal Quantum Computer -- a Quantum Computer has the same computational power as a TM, it is faster but its computational power is the same as a TM. In the same time in \cite{deutsch:reality} he says something that -- from a certain point of view -- is similar to a basic principle of hypercomputation:

\begin{quote}
The theory of computation has traditionally been studied almost entirely in the abstract,  as a topic in pure mathematics. This is to miss the point of it. Computers are physical objects, and computations are physical processes. What computers can or cannot compute is determined by the laws of physics alone, and not by pure mathematics.
\end{quote}

This quote can be perceived in two ways:
\begin{itemize}
\item From a hypercomputational point of view: One can create a new computer model as long as it does not conflict with any laws of physics.
\item From a classical point of view: It is useless to create a computational model based on pure mathematics. A computer must be physically realizable.\\
\end{itemize}

As mentioned before, there are quite a variety of Qauntum Computer models. In this chapter, after a brief introduction of some necessary concepts of quantum mechanics and quantum computing, we will explore a model that has generated some heated discussions in the scientific world: Tien D. Kieu's Adiabatic Quantum Computing. It is said that Kieu's Adiabatic Quantum Computer is able to answer \emph{Hilbert's $10$--th problem}, a problem which no classical computer can give an answer to. But first -- in order to better understand the model -- let's start with the basis.

\section{Qubit. What? Why? When?}

As we know, at the very core of any modern digital equipment are the bits. In order to make modern digital equipments work we must manipulate bits. A bit is like an abstract data type; there are many ways to represent it. The most crucial thing to have is a way of distinguishing between its two values: \textbf{0} and \textbf{1}. Once we can distinguish these values, store and manipulate them we can create all the digital machines we need. 

Today this is all given for granted: we can easily store, read and modify the value of a bit.

Richard Feynman, in his paper \cite{feynman:bottom}, alluded to the possibility of a further miniaturization of digital aparatuses and also anticipated that very small objects would be manipulated by the laws of quantum mechanics rather than classical mechanics. Given that the values of the bits must also be stored in some sort of physical support and that these supports would become smaller and smaller, for the explanation of their behaviour and for their manipulation quantum physics would be necessary. At that point, what we now know about bits and their manipulation will no longer be true.

There are quite some substantial differences between the quantum bits -- or, \emph{Qubits} -- and the classical bits we know today. For example:
\begin{itemize}
\item A classical bit can be in only one of its two possible states -- \textbf{0} or \textbf{1} -- while a qubit can be in a \emph{superposition} of these states.
\item A classical bit can be read or copied without altering its state or altering the states of other bits while a qubit can not be read or copied without altering its state and -- in case it is \emph{entangled} with other qubits -- without altering the states of the qubits it is entangled with.\\
\end{itemize}

Obviously, to use the qubit, one needs a way to manipulate them. To do so, one can proceed in two different directions:
\begin{enumerate}
\item One can try and suppress their quantistic ``side-effects'' and reduce the quantum system to a classical one, or
\item One can use these quantistic ``side-effects'' and try to create something new.\\
\end{enumerate}

Luckily, quantum systems possess some properties that help us in the encoding of their states in bits. For example, when we measure the \emph{spin} of an electron we find that it can have two possible values: \emph{spin-up} and \emph{spin-down}.\footnote{When the spin is parallel to the measurement axis it is called \emph{spin-up} and it is called \emph{spin-down} when it is not.} If a quantum system has two states it can be used to encode the values \textbf{0} and \textbf{1}. If the system used to represent the qubit is a quantum system it will be called \emph{qubit}.

\subsection{Representing a qubit.}

As we will see later on, in quantum mechanics, to each physical system is associated a proper vector space where an inner product is possible and where each vector represents a possible state of the system. Qubits are no exception, their states are representet via vectors in such a space. In quantum mechanics, instead of the standard geometric notation, is used a notation first introduced by the British physicist Paul Dirac. This notation is called \emph{Dirac Notation} or \emph{bra-ket notation}. The inner product in this notation is represented by a $\left\langle \textbf{bra} | c | \textbf{ket} \right\rangle$: \[ \left\langle \varphi | \psi \right\rangle \] made from a left side $\left\langle \varphi \right|$ called \emph{bra} and a right side $\left| \psi \right\rangle$ called \emph{ket}.

To better understand the Dirac Notation I find it useful to start from the standard concept of a vector in a three dimensional euclidean space where a vector $\vec{v}$ is a geometric entity endowed with magnitude and direction. As it is well known, if we have a Cartesian reference system consisting of the three axes $x$, $y$ and $z$ where $\vec{i}$, $\vec{j}$ and $\vec{k}$ are their relative unit vectors, each vector $\vec{v}$ can be expressed as a linear combination of these three unit vectors:
\[ \vec{v} = a\vec{i} + b\vec{j} + c\vec{k} \; \mathrm{where} \; a,b,c \in \mathbb{R} \]
In this way we can identify a vector by three numbers which can be represented as a column -- or row -- matrix. The three unit vectors are represented by the following matrices:
\[ \vec{i} = \begin{bmatrix} 1\\ 0\\ 0 \end{bmatrix} \; \vec{j} = \begin{bmatrix} 0\\ 1\\ 0 \end{bmatrix} \; \vec{k} = \begin{bmatrix} 0\\ 0\\ 1 \end{bmatrix} \]
and for a generic vector $\vec{v}$ we have:
\[ \vec{v} = a\vec{i} + b\vec{j} + c\vec{k} = a \begin{bmatrix} 1\\ 0\\ 0 \end{bmatrix} 	+ b \begin{bmatrix} 0\\ 1\\ 0 \end{bmatrix} + c \begin{bmatrix} 0\\ 0\\ 1 \end{bmatrix} = \begin{bmatrix} a\\ b\\ c \end{bmatrix} \]

Each row in the matrix represents a dimension and the multiplication factor of the unit vector which represents that dimension. Vector spaces in quantum mechanics are a simple generalization of the vector space concept in the Euclidean geometry, where:
\begin{enumerate}
\item The number of dimensions is not limited to only three, but it can be any number. In quantum mechanics the number of dimensions can even be infinite.
\item The multiplying factors are not limited to real numbers -- $n \in \mathbb{R}$ -- but are complex ones; in mathematical terms one can say that in quantum mechanics one deals with \emph{complex vector spaces} and not with real ones.
\end{enumerate}

Starting from the properties of the Euclidean space one can derive the axioms that define the notion of \emph{complex vector space} (which in quantum mechanics is represented with \emph{kets}):
\begin{enumerate}
\item $\left| \alpha \right\rangle + \left| \beta \right\rangle = \left| \beta \right\rangle + \left| \alpha \right\rangle$
\item $\left( \left| \alpha \right\rangle + \left| \beta \right\rangle \right) + \left| \gamma \right\rangle = \left| \alpha \right\rangle + \left( \left| \beta \right\rangle + \left| \gamma \right\rangle \right)$
\item $\exists \mathbf{0}, \left| \alpha \right\rangle + \mathbf{0} = \left| \alpha \right\rangle$ where $\mathbf{0}$ represents the vector with a length equal to $0$.
\item $\exists \left| -\alpha \right\rangle, \left| \alpha \right\rangle + \left| -\alpha \right\rangle  = \mathbf{0}$
\item $a \left( \left| \alpha \right\rangle + \left| \beta \right\rangle \right) = a \left| \alpha \right\rangle + a \left| \beta \right\rangle$
\item $\left( a + b \right) \left| \alpha \right\rangle = a \left| \alpha \right\rangle + b \left| \alpha \right\rangle$
\item $a \left( b \left| \alpha \right\rangle \right) = ab \left| \alpha \right\rangle$
\item $\left| -\alpha \right\rangle = -1 \left| \alpha \right\rangle$\\
\end{enumerate}

Each vector space can be associated in a one-to-one correspondence with a dual space. In the Dirac Notation a vector belonging to the dual space is represented by a \emph{bra}. In order to efficiently operate in the vector space using the rules with which one operates on matrices a ket is represented by a column matrix whilst a bra is represented by a row matrix. To each ket:
\[ \left| \psi \right\rangle = \begin{bmatrix} c_1 \\ c_2 \\ \vdots \\ c_n \end{bmatrix} \]
corresponds the bra:
\[ \left\langle \psi \right| = \begin{bmatrix} c_1^* & c_2^* & \cdots & c_n^* \end{bmatrix} \]
where $c_i^*$ is the complex conjugate of $c_i$. The inner product of the kets $\left| \psi \right\rangle$ and $\left| \varphi \right\rangle$ in the Dirac Notation is the moltiplication of the matrix that represents the bra corresponding to the first ket with the matrix that represents the second ket:
\[ \left| \psi \right\rangle \centerdot \left| \varphi \right\rangle = \left\langle \psi | \varphi \right\rangle \]
and it has the following properties:
\begin{enumerate}
\item $\left \langle \beta | c_1 \alpha_1 + c_2 \alpha_2 \right \rangle =
        c_1 \left \langle \beta | \alpha_1 \right \rangle + c_2 \left \langle \beta | \alpha_2 \right \rangle$
\item $\left \langle \beta | \alpha \right \rangle^* = \left \langle \alpha | \beta \right \rangle$
\item $\left \langle \alpha | \alpha \right \rangle \geq 0 \; e \; \left \langle \alpha | \alpha \right \rangle =
        0 \Leftrightarrow \left | \alpha \right \rangle = 0$\\
\end{enumerate}

The \emph{norm} of the vector:
\[ \left| \psi \right\rangle = \begin{bmatrix} c_1 \\ c_2 \\ \vdots \\ c_n \end{bmatrix} \]
is defined as:
\[
\| \psi \| = \sqrt{\left\langle \psi | \psi \right\rangle} = \sqrt{c_1^* c_1 + c_2^* c_2 + \cdots + c_n^* c_n} =
            \sqrt{|c_1|^2 + |c_2|^2 + \cdots + |c_n|^2}
\]
As one can see, the norm of a vector is a non negative real number, as one expects a length to be. In quantum mechanic the states of a system are represented by unit vectors:
\[ \| \psi \| = \sqrt{\left\langle \psi | \psi \right\rangle} = 1 \]
or rather:
\[ |c_1|^2 + |c_2|^2 + \cdots + |c_n|^2 = 1 \]
which in the case of a single qubit system is translated in:
\[ \| \psi \| = |a|^2 + |b|^2 = 1 \]

Using matrices, we have that the inner product of:
\[
\left| \psi \right\rangle = \begin{bmatrix} a \\ b \end{bmatrix}
\; \mathrm{with} \; 
\left| \varphi \right\rangle = \begin{bmatrix} c \\ d \end{bmatrix}
\]
is:
\[ \left\langle \psi | \varphi \right\rangle = \begin{bmatrix} a^* & b^* \end{bmatrix} \begin{bmatrix} c \\ d \end{bmatrix} = a^* c + b^* d \]

Another way to represent a qubit is the \emph{Bloch Sphere}, but for the purpose of this thesis it is not a helpful concept. For more information regarding the Dirac Notation or the Bloch Sphere one can read \cite{shankar:quantum} and \cite{williams:qc}.

\subsection{Properties of Qubits.}

In this section we'll give a brief explanation of some fundamental properties of qubits. As mentioned before, a qubit is very different from a classical bit and here are some of the reasons why:

\begin{enumerate}
\item \textbf{The state of a qubit is a vector.}

	As we saw before, a qubit is associated to an abstract two-dimensional vector space. Therefore the states of a base are two and are indicated -- by analogy with the classical bits -- with the kets $\left| 0 \right\rangle$ and $\left| 1 \right\rangle$ which -- from a physical point of view -- may correspond to the spin-up or spin-down of a particle. The state of a qubit is represented by a unit vector belonging to such space: 	\[ \left| \psi \right\rangle = a \left| 0 \right\rangle + b \left| 1 \right\rangle \] where $a$ and $b$ are complex numbers -- $a,b \in \mathbb{C}$ -- such that $||a||^2 + ||b||^2 = 1$. One must not confuse the state $\left| 0 \right\rangle$ with the vector $\mathbf{0}$ which does not represent a state.
	
\item \textbf{The amount of information obtainable from a qubit is the same as the amount of information obtainable from a classical bit.}

	The quantity of possible states of a qubit is infinite because such is the quantity of the possible linear combinations of its base states. The coefficients $a$ and $b$ are complex numbers of infinite precision so, apparently, the state of a qubit contains an infinite quantity of information, information which is ``hidden'', not accessible in any way. In fact, when one tries to make a measurement of the state of the qubit the result is reduced to one of the two possible states: $\left| 0 \right\rangle$ or $\left| 1 \right\rangle$. In other words, from the measurement of the state of a qubit one can have as a result the values:
	\begin{itemize}
	\item $0$ with a probability of $||a||^2$
	\item $1$ with a probability of $||b||^2$
	\end{itemize}
	Holevo's theorem -- \cite{wilde:qstheory}, section 11.6 -- gives further proof of the fact that the amount of information obtainable from a qubit is the same as the amount of information obtainable from a classical bit.
	
\item \textbf{Quantum entanglement.}

	A system of more than one qubit may exhibit the phenomenon known as \emph{Qauntum Entanglement}. The entanglement is a property which allows the state of a qubit to influence the state its entangled qubit. For example, a possible state $\left| \psi \right\rangle$ of two entangled qubits may be:
	\[ \left| \psi \right\rangle = \frac{1}{\sqrt{2}}\left| 0 0 \right\rangle + \frac{1}{\sqrt{2}}\left| 1 1 \right\rangle \]
	or
	\[ \left| \psi \right\rangle = \frac{1}{\sqrt{2}}\left| 0 1 \right\rangle + \frac{1}{\sqrt{2}}\left| 1 0 \right\rangle \]
	
	In the case a measurement should be made on the first qubit and its state is $\left| 0 \right\rangle$ one would know that the state of the second qubit will be $\left| 0 \right\rangle$ in the first case and $\left| 1 \right\rangle$.\\
\end{enumerate}

\section{Some helpful concepts.}

In this section we will briefly introduce -- as the title says -- some helpful concepts for a better understanding of the last part of this thesis.

\subsection{Fermat's last theorem.}

Fermat's last theorem asserts that there exist no three integers $a$, $b$, $c$ that satisfy the equation:
\[ a^n + b^n = c^n \]
for every $n > 2$. Fermat did not give any proof for all the numbers but only for one case: $n = 4$.

The equation $a^n + b^n = c^n$ is an example of a \emph{Diophantine equation}. A Diophantine equation is a polynomial equation the variables of which can only be integer numbers. While analyzing a Diophantine equation some of the questions that are usually asked are:
\begin{enumerate}
\item Are there any solutions?
\item If some solutions have been already found, can we find any more solutions?
\item Is there a finite or an infinite quantity of solutions?
\item Is it possible to find all the solutions?
\item Are the solutions of the equation computable?\\
\end{enumerate}

In the year 1900 Hilbert gave a list of 23 mathematical problems which were not solved until then. The computability of all the Diophantine equations was the $10$-th problem in that list.\footnote{That is why this problem is usually referred to as \emph{Hilbert's tenth problem}.} In the year 1970 Yuri Matiyasevich gave proof of the non-computability of this problem. Afterwards, Hilary Putnam and other authors gave proof that each recursively enumerable set was a \emph{Diophantine set}, result known as the \emph{Matiyasevich theorem}.

\subsection{Linear operators.}

A \emph{linear operator} in a vector space $V$ is a function $\Omega: V \rightarrow V$ with the following properties:
\begin{enumerate}
\item $\Omega (\alpha \left| V \right\rangle) = \alpha \Omega \left| V \right\rangle$
\item $(\left\langle V \right| \alpha) \Omega = \left\langle V \right| \Omega \alpha$
\item $\Omega (\alpha \left| V_i \right\rangle + \beta \left| V_j \right\rangle) = \alpha \Omega \left| V_i \right\rangle +
        \beta \Omega \left| V_j \right\rangle$
\item $(\left\langle V_i \right| \alpha + \left\langle V_j \right| \beta) \Omega = \left\langle V_i \right| \Omega \alpha + 
		\left\langle V_j \right| \Omega \beta$\\
\end{enumerate}
These operators are usually representated by matrices.

Sometimes, by applying a linear operator to a vector $\left| v \right\rangle$ we obtain a multiple of this vector, that is:
\[ \Omega \left| v \right\rangle = \omega \left| v \right\rangle \]
where:
\begin{itemize}
\item $\Omega$ is the linear operator
\item $\omega$ is a scalar.\\
\end{itemize}

An example of such a situation may be:
\[
\Omega \left| V \right\rangle = \begin{bmatrix} 0 & 3 \\ 3 & 0 \end{bmatrix} \begin{bmatrix} 2 \\ 2 \end{bmatrix} 
								= \begin{bmatrix} 0 \cdot 2 + 3 \cdot 2 \\ 3 \cdot 2 + 0 \cdot 2 \end{bmatrix} 
								= \begin{bmatrix} 6 \\ 6 \end{bmatrix} 
								= 3 \cdot \begin{bmatrix} 2 \\ 2 \end{bmatrix} = 3 \left| V \right\rangle
\]
In these cases $\left| v \right\rangle$ is an \emph{eigenvector} of $\Omega$ with an \emph{eigenvalue} equal to $\omega$.

Given a linear operator $\Omega$ one can give proof that there is only one operator $\Omega^{\dagger}$ such that:
\[ \left\langle \varphi | \Omega \psi \right\rangle = \left\langle \Omega^{\dagger} \varphi | \psi \right\rangle \]
$\Omega^{\dagger}$ is the \emph{adjoint operator} of $\Omega$.\footnote{In the case of matrices, the matrix $A^{\dagger}$ is the conjugate trasposed of the matrix $A$. For further concepts regarding matrices refer to Appendix \ref{matrici}.} An operator $\Omega$ which is equal to its adjoint operator -- $\Omega = \Omega^{\dagger}$ -- is a \emph{self-adjoint operator} or \emph{Hermitian operator}. In the case of a Hermitian operator we have that:
\[ \left\langle \varphi | \Omega \psi \right\rangle = \left\langle \Omega \varphi | \psi \right\rangle = \left\langle \varphi | \Omega | \psi \right\rangle \]

The eigenvalues of Hermitian operators are all real values. For each measurable physical quantity of a system there is a Hermitian operator in the vector space associated with that system and the eigenvalues of such operator represent the possible results returned by a measurement of the corresponding physical quantity.

The states that are eigenvectors of a Hermitian operator are also called \emph{eigenstates} of that operator. The eigenstates of a Hermitian operator corresponding to distinct eigenvalues are orthogonal between them.\footnote{If the inner product of two vectors $\left| \varphi \right\rangle$ and $\left| \psi \right\rangle$ is null -- $\left\langle \varphi | \psi \right\rangle = 0$ -- then these vectors are orthogonal between them.}

The measuring of a physical quantity makes the system go from the state $\left| \psi \right\rangle$ -- in which the system was right before the measuring operation -- to one of the states $\left| \varphi_i \right\rangle$ -- which are eigenvectors of the operator associated to that particular physical quantity -- with a probability of $|| \left\langle \varphi_i | \psi \right\rangle ||^2$. For example, if we measure the value of a qubit which is in a state $\left| \psi \right\rangle = a \left| 0 \right\rangle + b \left| 1 \right\rangle$ its state will become one of the following:
\begin{itemize} 
\item $\left| 0 \right\rangle$ with a probability of $|| \left\langle 0 | \psi \right\rangle ||^2 = ||a||^2$
\item $\left| 1 \right\rangle$ with a probability of $|| \left\langle 1 | \psi \right\rangle ||^2 = ||b||^2$
\end{itemize}
because $\left| 0 \right\rangle$ and $\left| 1 \right\rangle$ are orthogonal.

For more information regarding the subject one can consult \cite{lang:algebra}, \cite{shankar:quantum} and \cite{reed:physics}.

\subsection{Tensor Product.}

As we saw, a qubit can be described by a two-dimensional vector space. To a $n$ qubit system is associated a $2^n$ dimensional space which is the result of the \emph{tensor product} of the $n$ two-dimensional spaces of the respective qubits.

The tensor product -- indicated with the symbol $\otimes$ -- of two vector spaces $H_1^n$ and $H_2^m$ gives as a result the space $H_3^{n \times m}$:
\[ H_3 = H_1 \otimes H_2 \]
The quantum gates and the quantum operators of $H_3$ will be represented by square matrices with dimensions $n \cdot m \times n \cdot m$.

Let's make an example: To a 2-qubit system is associated a vector space of $2 \cdot 2 = 4$ dimensions. The quantum operators and the quantum gates of this system will be represented by $4 \times 4$ matrices. The matrix representing a quantum operator of the space $H_3$ -- let's say, $\Omega_{H_3}$ -- will be the result of the tensor product between the matrix representing $\Omega_{H_1}$ and the matrix representing $\Omega_{H_2}$:
\[ \Omega_{H_3} = \Omega_{H_1} \otimes \Omega_{H_2} \]
Let us suppose that $\Omega$ is the \textbf{NOT} operator. In this case we would have that:
\[
\mathbf{NOT}_{H_3} = \mathbf{NOT}_{H_1} \otimes \mathbf{NOT}_{H_2} = \begin{bmatrix} 
	0 & 1 \\ 
	1 & 0 
\end{bmatrix} \otimes \begin{bmatrix}
	0 & 1 \\
	1 & 0
\end{bmatrix} = \begin{bmatrix}
	0 & 0 & 0 & 1 \\
	0 & 0 & 1 & 0 \\
	0 & 1 & 0 & 0 \\
	1 & 0 & 0 & 0
\end{bmatrix}
\]

One can find a brief explanation of the tensor product of matrices in Appendix B of the present work.

\subsection{Fock space.}

The \emph{Fock space} is defined as the resulting vector space $H$ of the sum of the tensor product of the vector spaces associated to single particle systems:
\[
F_v(H) = \bigoplus_{i = 0}^{\infty} S_v H^{\otimes n} = \mathbb{C} \oplus H \oplus \left( S_v \left(H \otimes H \right) \right) \oplus 
		\left(S_v \left(H \otimes H \otimes H \right) \right) \oplus \dots
\]
where:
\begin{itemize}
\item $S_v$ is the operator which symmetrizes or antisymmetrizes a tensor.\footnote{Depending on the type of particle.}
\item $\mathbb{C}$ represents the states of no particles.
\item $H$ represents the state of one particle.
\item $S_v(H \otimes \cdots \otimes H)$ represents the states of $n$ particles.
\end{itemize}

A generic state $\left| \Psi \right\rangle$ in $F_v{H}$ is given by:
\[ \left| \Psi \right\rangle = \psi_0 \oplus \left| \psi_1 \right\rangle \oplus \left| \psi_{11}, \psi_{12} \right\rangle \oplus \dots \]
where:
\begin{itemize}
\item $\psi_0$ is a complex number.
\item $\left| \psi_1 \right\rangle \in H$
\item $\left| \psi_{11}, \psi_{12} \right\rangle \in S_v(H \otimes H)$, etc.
\end{itemize}

The inner product $\left\langle \Psi | \Phi \right\rangle_v$ in $F_v(H)$ is defined as:
\[
\left\langle \Psi | \Phi \right\rangle_v =
    \psi_0^* \phi_0 + \left\langle \psi_1 | \phi_1 \right\rangle + \left\langle \psi_{11}, \psi_{12}| \phi_{11}, \phi_{12} \right\rangle +
    \dots
\]
where the inner products on each of the $n$-particle Hilbert spaces are used. The basis of a Fock space is made of the \emph{Fock states}, which can be described as elements of a Fock space with a well-defined number of particles. For a more in-depth explanation of the Fock spaces one can consult \cite{reed:physics}.

\subsection{The Schr\"odinger equation and the Hamiltonian operator.}

The evolution over time of a closed quantum system -- the one which occurs when the system, after its initial preparation, is retained isolated from the external environment and is not subjected to measurement -- is given by a differential equation, the fundamental equation of quantum mechanics, the \emph{Schr\"odinger equation}:
\[ i\hbar\frac{\partial \left | \psi (t) \right \rangle}{\partial t} = \hat{H} \left | \psi (t) \right \rangle \]
where $\hbar$ is the \emph{Planck constant} and $\hat{H}$ is the \emph{Hamiltonian} of the system. The Hamiltonian -- or \emph{Hamiltonian operator} -- is a hermitian operator which corresponds to the total energy of the system. The eigenvalues of the Hamiltonian are the possible values of the energy of the system.

As one can notice, given the fact that $\hbar$ and $i$ are constants and having the state $\left| \psi(t_0) \right\rangle$ of the system in the initial time $t_0$, the evolution of the system in time -- the state $\left| \psi(t) \right\rangle$ of each following instant -- is uniquely determined by the Hamiltonian $\hat{H}$. As the laws of Newton for classical mechanics, the Hamiltonian allows us -- given the initial conditions -- to predict the behaviour of an isolated dynamic system.

An isolated system that is in an eigenstate of the Hamiltonian stays in that over time. For this reason the eigenstates of the Hamiltonian are also called \emph{stationary states}.

In most cases, even the evolution over time of a not isolated system can be described by the Schr\"odinger equation. In this case the Hamiltonian -- instead of being a determined operator for that system -- will be an operator $\hat{H}(t)$ which varies with time.

A more exhaustive explanation of the topic can be found in \cite{williams:qc} and \cite{shankar:quantum}.

\subsection{Adiabatic Theorem.}

The phrase \emph{Adiabatic Process} is used in thermodynamics to indicate those processes during which there is no heat exchange between the system and the environment which surrounds it. This happens when the system evolves much faster than the surrounding environment. In quantum mechanics the phrase is used to indicate processes during which the Hamiltonian $\hat{H}(t)$ of the system varies very slowly, infinitely slowly.

One should notice that if an operator is time-dependent its eigenstates and eigenvalues will also be time-dependent. If a system with a time-dependent Hamiltonian $\hat{H}(t)$ is in an eigenstate $\left| \psi(t_0) \right\rangle$ of its Hamiltonian -- in an initial time indicated with $t_0$ -- $\hat{H}(t_0)$ we can ask in what state $\left| \psi(t_1) \right\rangle$ will the system be in a following time $t_1$ and if that state will still be an eigenstate of the Hamiltonian $\hat{H}(t_1)$. Usually this does not occur, unless the system evolves according to an adiabatic process.

According to the Adiabatic Theorem, a system which -- in a point-in-time $t_0$ -- is in an eigenstate $\left| \psi(t_0) \right\rangle$ of the Hamiltonian $\hat{H}(t_0)$ with an eigenvalue $E(t_0)$ will be -- in a point-in-time $t_1$ -- in the corresponding eigenstate $\left| \psi(t_1) \right\rangle$ of $\hat{H}(t_1)$ \emph{if the Hamiltonian varies slowly enough}, that is if $\frac{dH}{dt}$ is small enough and if \emph{the initially distinct eigenvalues of $\hat{H}$ remain that way}. For the theorem to be true it is also required that the first and second derivates of the instantaneous eigenvectors with respect to time must be well defined and piecewise continuous. 

\begin{quote}
\dots if we take a quantum system whose Hamiltonian slowly changes from $H_1$ to $H_2$, then, under certain conditions on $H_1$ and $H_2$, the ground (lowest energy) state of $H_1$ gets transformed to the ground state of $H_2$. \cite{ambainis:adiabatic}
\end{quote}

A more in-depth explanation of the topic can be found in \cite{kieu:computing}, \cite{amin:adiabatic} and \cite{ambainis:adiabatic}.

\section{Adiabatic Quantum Computing.}

In this section we will talk about a hypercomputational quantum computer. As we will see later, this is a probabilistic computer, given the fact that it is based on the adiabatic theorem. Like anything new -- quite rightly, i would dare say -- it has raised many criticisms and objections in the scientific world, some of which will be seen further ahead in the present work.

For the introduction of the model we will take as reference the work of the creator of the model himself, in particular to his papers: \cite{kieu:aqc} and \cite{kieu:computing}.

\subsection{The Adiabatic Quantum Computer.}

The \emph{Adiabatic Quantum Computer (AQC)} was introduced for the first time in \cite{farhi:adiabatic} and then it was resubmitted, with some differences, by Kieu. The idea behind the model is to encode the solution of a problem in the \emph{ground state} $\left| g_F \right\rangle$ of an appropriate Hamiltonian $H_F$. However, since this state is very difficult to achieve, the system is prepared in a more feasible ground state $\left| g_I \right\rangle$ of another Hamiltonian $H_I$ and then this Hamiltonian is slowly turned in the wanted Hamiltonian $H_F$, according to the formula:
\[ H\left(\frac{t}{T}\right) = \left( 1 -- \frac{t}{T} \right) H_I + \frac{t}{T} H_F \]
where $T$ is the time needed for the transformation of $H_I$ in $H_F$.

According to the adiabatic theorem, if the time $T$ used to transform $H_I$ in $H_F$ compared to the inner time scale of the system is long enough, it is very likely that the initial state will evolve in the desired state.\footnote{Saying it differently: The greater the time available, the greater the probability for the transformation to be successful.}

\subsection{The solution to the Hilbert's tenth problem.}

Let us consider the following Diophantine equation:
\begin{equation} \label{aqc:eq1}
\left( x + 1 \right)^3 + \left( y + 1 \right)^3 + \left( z + 1 \right)^3 + cxyz = 0
\end{equation}
where:
\begin{itemize}
\item $c \in \mathbb{Z}$
\item $x, y, z$ \emph{unknown}
\end{itemize}
Is it possible to know if the equation has any integer solution? In \cite{kieu:computing} the author introduces a decision algorithm which, given a Diophantine equation, should be able to answer the before-mentioned question.

The author states that to have a solution at the before-mentioned problem one must implement a Fock space. In that space must be created the Hamiltonian which corresponds to (\ref{aqc:eq1}):
\begin{equation} \label{aqc:hamiltonian}
H_P = \left(
        \left( a_x^{\dagger} a_x + 1 \right)^3 +
        \left( a_y^{\dagger} a_y + 1 \right)^3 +
        \left( a_z^{\dagger} a_z + 1 \right)^3 +
        c\left( a_x^{\dagger} a_x \right)\left( a_y^{\dagger} a_y \right)\left( a_z^{\dagger} a_z \right)
    \right) ^2
\end{equation}

The operators $N_j = a_j^{\dagger}a$ have non-negative integer eigenvalues $n_j$. The ground state $\left| g \right\rangle$ of $H_P$ has the following properties:
\begin{itemize}
\item $N_j \left| g \right\rangle = n_j \left| g \right\rangle$
\item $H_P \left| g \right\rangle = \left(
		\left( n_x + 1 \right)^3 +
		\left( n_y + 1 \right)^3 +
		\left( n_z + 1 \right)^3 +
		cn_xn_yn_z
	\right)^2 \left| g \right\rangle \equiv E_g \left| g \right\rangle$ for some $(n_x, n_y, n_z)$\\
\end{itemize}

The answer to the decision problem will be given by the projective measurement of the energy $E_g$ of the state $\left| g \right\rangle$. If $E_g = 0$ then the Diophantine equation has at least one integer solution, otherwise it does not have any integer solutions.

The algorithm can be summarized in:
\begin{enumerate}
\item Implement the Hamiltonian 
	\[ H_P = \left( D \left( a_1^{\dagger}a_1, \dots , a_n^{\dagger}a_n \right) \right)^2 \]
	corresponding to a Diophantine equation with $n$ unknowns
	\[ D \left( x_1, \dots , x_n \right) = 0 \]
	in an appropriate Fock space.
	
\item If the ground state can be obtained with a high probability and can be virified, the measurement of some observables will give the answer to the decision problem.\\
\end{enumerate}

For more information regarding the topic one can consult \cite{kieu:computing}.

\subsection{Critics to the model.}

There have been quite a few critics to Kieu's Adiabatic Quantum Computing model in the scientific community. These critics can be roughly divided in two groups:
\begin{itemize}
\item Critics to the algorithm.
\item Critics to its feasibility.\\
\end{itemize}

The objections regarding its feasibility refer to the fact that the model is based on an infinite Fock space. The Hamiltonians of these spaces can not be built as it is not possible to make measurements with infinite precision. Regarding the other category of objections -- the critics to the algorithm -- there is still a debate on its correctness going on.

This model is in contrast with what was claimed by D. Deutsch regarding the computing power of a quantum computer. According to Deutsch, a quantum computer calculates the same class of function of a TM, even though it does it much more efficiently and much more rapidly.\footnote{A quantum computer can calculate, in a reasonable time, functions that are intractable for a TM.} This model, unlike the model presented by David Deutsch in \cite{deutsch:qtm}, is based on Hamiltonians of infinite dimensions -- which operate in a Fock space -- and the properties of their ground states.

In \cite{aharonov:aqc} the authors demonstrate that the AQC is equivalent -- in terms of calculating power -- to the standard quantum computing.\footnote{To the model presented by Deutsch.} This demonstration, however, is only valid for Hilbert spaces of finite dimensions. The hypercomputability of the AQC is based on its probabilistic nature and on the fact that it works on spaces of infinite dimensions.

In \cite{hagar:quantum} the authors point out another problem:
\begin{quote}
[ \dots ] the fact that the global minimum for the ‘computed’ function exists by construction (which ensures a non–zero energy gap and hence a finite evolution time) is of no consequence. Rather, it is the fact that this finite time is unbounded which kills the algorithm. And since Kieu, while guaranteeing that the brute–force search will eventually halt, fails to supply a criterion that would allow one to identify whether or not the algorithm has halted on the global minimum, the whole construction, despite his aspirations, lacks the ability to identify a global minimum as such.The problem is thus no different than any other corresponding classical case of undecidability, and quantum mechanics adds nothing to its solution. \cite{hagar:quantum}
\end{quote}

\begin{quote}
Put another way, the gist behind the adiabatic algorithm is that after a sufficiently long evolution time, one is certain to have retrieved the correct result of the decision problem just by performing a measurement on the ground state. However, when the evolution time is unknown, a non–zero energy reading upon a measurement of a final state can be interpreted in two very different ways. On one hand, it may be said to be an eigenvalue of an excited state. In such case, clearly, the evolution was non–adiabatic, hence one must iterate the algorithm with another, longer, evolution time. On the other hand, it may be said to be an eigenvalue of the ground state. In such case, clearly, the algorithm has performed correctly and one has a (negative) answer to the decision problem. But since one cannot check a negative answer to a classically undecidable problem, how can one tell, without knowing $T$ in advance, that this negative ‘answer’ is indeed correct, that is, that no iterations are needed anymore? Without a criterion for distinguishing a ground state from all other excited states which is independent of the knowledge of the adiabatic evolution time $T$, one simply can’t. \cite{hagar:quantum}
\end{quote}

The fundamental problem is to be able to determine if the answer given by the machine is correct. In fact -- like in the Halting problem -- if the answer is positive, if the Diophantine equation has a solution, it is easy to verify. If, on the other hand, the answer is negative, one can not be sure of the correctness of that answer. On this aspect Kieu replies:
\begin{quote}
The fact that our algorithm is ``only'' probabilistically correct can be understood as a necessity and a consistency condition when the outcomes of such an algorithm cannot, in principle, be verified by any other means. The algorithm gives the k-tuple at which the square of a Diophantine polynomial assumes it smallest value. While the existence of a solution can be verified by a simple substitution, the indication of no solution cannot be verified by any other finite recursive means at all – thus the need of some probability measure to quantify the accuracy of the derived conclusion. However, it is important and useful that this probability is not only known but can also be predetermined with an arbitrary value in advance. \cite{kieu:hypercomputability}
\end{quote}

For more information regarding the critics made to the model one can read \cite{aharonov:aqc}, \cite{kieu:hypercomputability}, \cite{kieu:hodges}, \cite{kieu:smith},  \cite{hagar:quantum} and \cite{smith:aqc}.

\chapter{Conclusions.}

In this thesis we have explored the key aspects of a relatively new field of study, the field of \emph{Hypercomputation}. We started with a thorough study and analysis of the classical concept of computability and of the Church-Turing thesis to then continue with a study of the concept of hypercomputation, the introduction of some hypercomputational machines and a more thoroughly analysis of some of those machines. We tried to focus on the aspect of their feasibility as well as on their computational power.

As one can guess from the machines introduced in this thesis, hypercomputational machines can be divided in two main groups:
\begin{enumerate}
\item Machines which require the use of the mathematics of the continuum to be described; \label{hyper:continuum}
\item Machines which extend the behaviour of TMs accepting a discrete description. \label{hyper:discrete}\\
\end{enumerate}

Regarding the machines belonging to the first group there is a fundamental problem, measuring with infinite precision.\footnote{See Scarpellini's quote in section \ref{hypercomputation:intro}.}

The machines belonging to the second group need to work on conditions different from those of a TM. Let's make a brief recapitulation of the conditions on which a TM works, described in more detail in \ref{hypercomputation:intro}:
\begin{enumerate}
\item A TM accepts input data only before the computation starts.	\label{tm:input}
\item A TM follows a set of fixed rules during its computation. \label{tm:rules}
\item If the computation of a TM returns a result, that result is obtained after a finite amount of time and is unique. \label{tm:output}\\
\end{enumerate}

Among these models exist some which are able to compute Supertasks, which are -- roughly speaking -- an infinite amount of operations in a finite amount of time. Between the model introduced in this thesis there are two which can compute these tasks:
\begin{itemize}
\item Accelerated Turing Machines (ATM).
\item Malament-Hogarth Machines (MHM).\\
\end{itemize}

Obviously, every TM is subject to the laws of physics and for each of its computations there are some limits over which the machine can not go beyond.\footnote{Some of these limits are presented in section \ref{tm:steps}.} These limits indirectly set a time limit for each step of the computation of a TM. Nevertheless, ATMs violate this limit. The study of these machines is interesting as it gives us an idea of what is possible to calculate if one has the possibility to compute supertasks.

As for the MHMs, according to modern physics, they do not violate any physical limit and are perfectly feasible. The problem with these machines, as indicated also in \ref{hypercomputation:mhm}, is purely practical: finding two suitable space-time lines. A certain kind of black holes have the necessary characteristics to allow the MHM's computation but even there some problems -- practical and theoretical -- arise.

The other hypercomputational machines introduced in this thesis violate at least one of the before-mentioned conditions. If we violate the \ref{tm:input} condition we can obtain the \emph{Coupled Turing Machines} (CTM). This machine is hypercomputational because it takes as input a continuous stream of data which is potentially random. This randomness can make this machine not usable from a practical point-of-view. What use does a machine for which is not known what it calculates have?

\emph{Trial-And-Error} (TAE) machines violate the \ref{tm:output} condition as this machine gives off an output before it computation has reached an end -- if it ever reaches an end -- and may not be unique as the machine may ``change its mind'' after some time. From a certain point-in-time the TAE machine will not ``change its mind'' anymore, but this point-in-time is not known a priori. The feasibility of these machines is not the real problem in this case, these machines do not violate any physical law. If we are satisfied with a ``possibly correct'' answer and if we are willing to accept a ``more correct'' last-minute answer then these machines are a viable alternative.

After the analysis of these hypercomputational machines we entered the complex and counterintuitive realm of quantum computing where we saw the \emph{Adiabatic Quantum Computer} (AQC). As we mentioned previously, a Canadian company has built the first commercial quantum computer based on this model. There are still some debates on the truthfulness of the statements of the company, that is if the computer is a real quantum computer or not. In \cite{kieu:computing} the author presented an algorithm which could solve the Hilbert's tenth problem using the AQC. This algorithm aroused quite some interest in the scientific world and also a lot of critics. This model is a probabilistic one and as such is able to give a partially correct answer. Even though it can be arranged for the answer to be more or less correct -- depending on the conditions, we can have a correct answer with an arbitrary high percentage of confidence -- it can never be $100\%$ correct. In this model is required to work in infinite Fock spaces and this fact has attracted quite some critics. As if that were not enough, the probabilistic nature of the model was not of any help, on the contrary it gave rise to quite some debates.

Luckily, hypercomputation is a new field and surely many more ideas have yet to be brought on the surface. We have also seen a few models of the human mind based on some hypercomputers. Up until now it seems that to hypercompute we have to give off the certainty of a correct result and be satisfied with a ``probably correct'' result. Will it be possible to do more than this?

\appendix
\chapter{Complex Numbers.}

Complex numbers are an extension of real numbers. They are formed by two parts:
\begin{itemize}
\item Real
\item Imaginary
\end{itemize}

A complex number $c \in \mathbb{C}$ is represented as:
\[ c = a + b \cdot i \]
where:
\begin{itemize}
\item $a,b \in \mathbb{R}$
\item $i = \sqrt{-1}$ is the imaginary unit
\end{itemize}

Complex numbers can be also seen as an ordered pair of real numbers $(a, b)$. In fact, complex numbers are in a two-way correspondece with the points of a plane, also known as the \emph{Complex Plane}. This plane is made of the real axis and orthogonal imaginary axis.

The complex numbers have the following properties:
\begin{enumerate}
\item $c + c' = (a,b) + (a',b') = (a + a', b + b')$
\item $c \cdot c' = (a,b) \cdot (a', b') = (aa' - bb', ab' + ba')$
\item $|c| = \sqrt{a^2 + b^2}$
\item The \emph{distance} between two points in the complex plane is calculated by the function: $d(c, c') = |c - c'|$
\item The \emph{complex conjugate} of $c = a + bi$ is $c^* = a - bi$ and has the following properties:
	\begin{enumerate}
	\item $(c_1 + c_2)^* = c_1^* + c_2^*$
	\item $(c_1 c_2)^* = c_1^* c_2^*$
	\item $(c_1 c_2^{-1})^* = (c_1 / c_2)^* = c_1^* / c_2^*$
	\item $(c^*)^* = c$
	\item $c^* = c \Leftrightarrow c \in \mathbb{R}$
	\item $|c^*| = |c|$
	\item $|c|^2 = c^* c$
	\end{enumerate}
\item $c^{-1} = |c|^{-2} c^* = c^* / |c|^2$
\item It is not possible to define an order for complex numbers. We can not have $c_1 \leq c_2 \Rightarrow c_1 + c_3 \leq c_2 + c_3$ for the complex numbers as we have for the real ones.
\item $\mathbb{C}$ is, at the same time, a \emph{one-dimensional complex vector space} and a \emph{two-dimensional real vector space}. In addition to being a real vector space it is also a \emph{normed vector space}\footnote{A vector space where each vector has a defined norm, or length.} and a \emph{complete metric space}.\footnote{A complete metric space is a metric space where every sequence of the form $\{ x_i | \forall \epsilon > 0 \exists x_j \; such \; that \; d(x_i, x_j) < \epsilon \}$ converge to an element of the space.\cite{reed:physics}}. As such, it is also a \emph{Hilbert space}.
\end{enumerate}

\chapter{Matrixes.} \label{matrix}

As it is well known, a matrix is a table of elements made of $n$ rows and $m$ columns. A matrix $A_{(n,m)}$ has the following form:
\[
A_{(n,m)} = \begin{bmatrix}
a_{1,1} & a_{1,2} & \cdots & a_{1,m} \\
a_{2,1} & a_{2,2} & \cdots & a_{2,m} \\
\vdots & \vdots & \ddots & \vdots \\
a_{n,1} & a_{n,2} & \cdots & a_{n,m}
\end{bmatrix}
\]

The element indicated as $a_{ij}$ is the element that is found in the row $i$ and the column $j$. Matrices can be used to represent vectors. In fact, vectors can be considered as simple matrices with a single row or with a single column. A matrix $1 \times m$ is called \emph{row matrix} whereas a matrix $n \times 1$ is called \emph{column matrix}.

Different operations are possible between matrices:
\begin{enumerate}
\item \textbf{Sum of matrices.} The sum between matrices -- let us say matrix $A$ and matrix $B$ -- is possible only between matrices with the same number of rows $n$ and the same number of columns $m$. The resulting matrix $C$ will also have $n$ rows and $m$ columns. Each element $c_{i,j}$ will be the result of the sum of the elements $a_{i,j}$ and $b_{i,j}$:
		\[
		C = A + B = \begin{bmatrix}
		a_{1,1} & a_{1,2} & \cdots & a_{1,m} \\
		a_{2,1} & a_{2,2} & \cdots & a_{2,m} \\
		\vdots & \vdots & \ddots & \vdots \\
		a_{n,1} & a_{n,2} & \cdots & a_{n,m}
		\end{bmatrix} + \begin{bmatrix}
		b_{1,1} & b_{1,2} & \cdots & b_{1,m} \\
		b_{2,1} & b_{2,2} & \cdots & b_{2,m} \\
		\vdots & \vdots & \ddots & \vdots \\
		b_{n,1} & b_{n,2} & \cdots & b_{n,m}
		\end{bmatrix}
		\]
		\[
		C = \begin{bmatrix}
		a_{1,1} + b_{1,1} & a_{1,2} + b_{1,2} & \cdots & a_{1,m} + b_{1,m} \\
		a_{2,1} + b_{2,1} & a_{2,2} + b_{2,2} & \cdots & a_{2,m} + b_{2,m} \\
		\vdots & \vdots & \ddots & \vdots \\
		a_{n,1} + b_{n,1} & a_{n,2} + b_{n,2} & \cdots & a_{n,m} + b_{n,m}
		\end{bmatrix}
		\]	
\item \textbf{Direct sum between matrices.} The result of the direct sum of the matrix $A_{(n,m)}$ with the matrix $B_{(p,q)}$ is the matrix $C_{(n+p, m+q)}$ made in the following way:
		\[
		C = A \oplus B = \begin{bmatrix}
		A & 0 \\ 0 & B
		\end{bmatrix}
		\]
	
	In general we have that:
	\[
	\bigoplus_{i = 1}^{n} A_i = \begin{bmatrix}
	A_1 & 0 & \cdots & 0 \\
	0 & A_2 & \cdots & 0 \\
	\vdots & \vdots & \ddots & \vdots \\
	0 & 0 & \cdots & A_n
	\end{bmatrix}
	\]
\item \textbf{Multiplication.} In the case of multiplication we may have two cases:
	\begin{enumerate}
	\item \textbf{Multiplication with a scalar.} If $B_{(n,m)}$ is the result of the multiplication of the matrix $A_{(n,m)}$ with a scalar $k$ then the element $b_{i,j}$ will be equal to the multiplication of the element $a_{i,j}$ with $k$:
			\[
			B = k \cdot A = k \begin{bmatrix}
			a_{1,1} & a_{1,2} & \cdots & a_{1,m} \\
			a_{2,1} & a_{2,2} & \cdots & a_{2,m} \\
			\vdots & \vdots & \ddots & \vdots \\
			a_{n,1} & a_{n,2} & \cdots & a_{n,m}
			\end{bmatrix} = \begin{bmatrix}
			ka_{1,1} & ka_{1,2} & \cdots & ka_{1,m} \\
			ka_{2,1} & ka_{2,2} & \cdots & ka_{2,m} \\
			\vdots & \vdots & \ddots & \vdots \\
			ka_{n,1} & ka_{n,2} & \cdots & ka_{n,m}
			\end{bmatrix}
			\]	
	\item \textbf{Multiplication between matrices.} The multiplication of a matrix $A_{(n,p)}$ with a matrix $B_{(p,m)}$ results in the matrix $C_{(n,m)}$.\footnote{The resulting matrix will have the number of rows of the first matrix and the number of columns of the second matrix. The number of columns of the first matrix must be equal to the number of rows of the second matrix.} Each element $c_{i,j}$ will be equal to the sum of the product of each element in the row $i$ of the matrix $A$ with each element of the column $j$ of the matrix $B$. In other words:
			\[ c_{i,j} = \sum_{l = 0}^{n-1} a_{i,l}b_{l,j} = a_{i,1}b_{1,j} + a_{i,2}b_{2,j} + \cdots + a_{i,n}b_{n,j} \]
		
		This case includes also the multiplication of a matrix $A_{(n,m)}$ with a vector $\vec{v}$ represented by a column matrix $m \times 1$. In this case, the resulting vector $\vec{r}$ will have the components:
		\[ r_i = \sum_{j = 1}^{m} a_{i,j}v_j = a_{i,1}v_1 + a_{i,2}v_2 + \cdots + a_{i,m}v_m \]
	\item \textbf{Matrix product (in terms of inner product).} 
		\[
		C = A \cdot B = \begin{bmatrix}
		a_{1,1} & a_{1,2} & \cdots & a_{1,m} \\
		a_{2,1} & a_{2,2} & \cdots & a_{2,m} \\
		\vdots & \vdots & \ddots & \vdots \\
		a_{n,1} & a_{n,2} & \cdots & a_{n,m}
		\end{bmatrix} \cdot \begin{bmatrix}
		b_{1,1} & b_{1,2} & \cdots & b_{1,m} \\
		b_{2,1} & b_{2,2} & \cdots & b_{2,m} \\
		\vdots & \vdots & \ddots & \vdots \\
		b_{n,1} & b_{n,2} & \cdots & b_{n,m}
		\end{bmatrix}
		\]
		\[
		C = \begin{bmatrix}
		a_{1,1}b_{1,1} & a_{1,2}b_{1,2} & \cdots & a_{1,m}b_{1,m} \\
		a_{2,1}b_{2,1} & a_{2,2}b_{2,2} & \cdots & a_{2,m}b_{2,m} \\
		\vdots & \vdots & \ddots & \vdots \\
		a_{n,1}b_{n,1} & a_{n,2}b_{n,2} & \cdots & a_{n,m}b_{n,m}
		\end{bmatrix}
		\]
	\item \textbf{Matrix tensor product.} This product is indicated with the symbol $\otimes$ and the resulting matrix $C_{(n \times k, m \times p)}$ of $A_{(n,m)} \otimes B_{(k,p)}$ is:
		\[
		C = A \otimes B = \begin{bmatrix}
		a_{1,1}B & a_{1,2}B & \cdots & a_{1,m}B \\
		a_{2,1}B & a_{2,2}B & \cdots & a_{2,m}B \\
		\vdots & \vdots & \ddots & \vdots \\
		a_{n,1}B & a_{n,2}B & \cdots & a_{n,m}B
		\end{bmatrix}
		\]
	\end{enumerate}
	
\item \textbf{Matrix transposition.} The transposed of matrix $A_{(n,m)}$ is the matrix $A_{(m,n)}^T$ where the element $a_{i,j}$ of $A^T$ is equal to the element $a_{j,i}$ of $A$:
	\[
	A^T = \begin{bmatrix}
	a_{1,1} & a_{1,2} & \cdots & a_{1,m} \\
	a_{2,1} & a_{2,2} & \cdots & a_{2,m} \\
	\vdots & \vdots & \ddots & \vdots \\
	a_{n,1} & a_{n,2} & \cdots & a_{n,m}
	\end{bmatrix}^T = \begin{bmatrix}
	a_{1,1} & a_{2,1} & \cdots & a_{m,1} \\
	a_{1,2} & a_{2,2} & \cdots & a_{m,2} \\
	\vdots & \vdots & \ddots & \vdots \\
	a_{1,n} & a_{2,n} & \cdots & a_{m,n}
	\end{bmatrix}
	\]
	
	For example, let us consider the matrix:
	\[
	A = \begin{bmatrix}
	4 & 7 & 1 \\
	0 & 5 & 3
	\end{bmatrix}
	\]
	
	Its transposed will be:
	\[
	A^T = \begin{bmatrix}
	4 & 0 \\
	7 & 5 \\
	1 & 3
	\end{bmatrix}
	\]
	
	In the case of a matrix made of complex numbers -- a complex matrix -- we can talk about its transposed conjugate. The transposed conjugate of a complex matrix $A$ is indicated with $A^{\dagger}$ and is build by transposing $A$ and replacing each element with its complex conjugate. If $A$ is equal to $A^{\dagger}$ then we have a \emph{Hermitian matrix}. An example of a Hermitian matrix would be:
	\[
	A = A^{\dagger} = (A^*)^T = \begin{bmatrix}
	1 & 2 + i & -i \\
	2 - i & 5 & -3+7i \\
	i & -3-7i & 0
	\end{bmatrix}
	\]
	
	Some of the properties of these matrices are:
	\begin{enumerate}
	\item $(A^{\dagger})^{\dagger} = A$
	\item $(A+B)^{\dagger} = A^{\dagger} + B^{\dagger}$
	\item $(cA)^{\dagger} = c^* \cdot A^{\dagger}$ dove $c \in \mathbb{C}$
	\item $(A \cdot B \cdot \cdots)^{\dagger} = \cdots \cdot B^{\dagger} \cdot A^{\dagger}$
	\end{enumerate}
\end{enumerate}

For more information regarding the topic one may consult \cite{lang:algebra}.

\backmatter

\chapter{Acknowledgements.}

The people to thank for the present thesis are way too many. I have to heartfully thank \emph{prof. Felice Cardone} and \emph{prof. Elio Giovannetti} for the patience showed in helping me to better understand some of the introduced concepts and for helping me to always ``keep the eyes on the ball'', and I must say that this was the hardest task of all.

A big ``Thank You!'' goes to \emph{Apostolos Syropoulos} and \emph{Tien D\~{u}ng Kieu} which have helped me to better understand some key concepts in hypercomputation and Adiabatic Quantum Computing.

I also have to thank my friends which, indirectly, have helped me keep my feet on the ground. Also a big ``Thank You'' goes to my colleagues for the support offered.

Many people have contributed -- directly or indirectly -- on the realization of this work and I can never thank them enough. For now I can just say:\\ \begin{center} THANK YOU! \end{center}


\end{document}